\documentclass[]{RAA}
\usepackage{graphicx}
\usepackage{natbib}
\providecommand{\bm}[1]{\mbox{\boldmath$#1$\unboldmath}}

\begin{document}
   \title{Cusp-core problem and strong gravitational lensing}
\volnopage{ {\bf 2009} Vol.\ {\bf 9} No. {\bf XX}, 000--000}
   \setcounter{page}{1}
   \author{Nan Li
   \and Da-Ming Chen}
\institute{National Astronomical Observatories, Chinese Academy of Sciences,
             Beijing 100012, China; {\it uranus@bao.ac.cn; cdm@bao.ac.cn}\\
\vs \no
   {\small Received [2009] [May] [12]; accepted [2009] [May] [18] }
}
\abstract{Cosmological numerical simulations of galaxy formation have led to the cuspy density profile of a pure cold dark matter  halo toward the center, which is in sharp contradiction with the observations of the rotation curves of cold dark matter-dominated dwarf and low surface brightness disk galaxies, with the latter tending to favor mass profiles with a flat central core. Many efforts have been devoted to resolve this cusp-core problem in recent years, among them, baryon-cold dark matter interactions  are considered to be the main physical mechanisms  erasing the cold dark matter (CDM) cusp into a flat core in the centers of all CDM halos. Clearly,  baryon-cold dark matter interactions are not customized only for CDM-dominated disk galaxies, but for all types, including giant ellipticals. We first fit the most recent high resolution observations of rotation curves with the Burkert profile, then use the constrained core size-halo mass relation to calculate the lensing frequency, and compare the predicted results with strong lensing observations. Unfortunately, it turns out that the core size constrained from rotation curves of disk galaxies cannot be extrapolated to giant ellipticals. We conclude that, in the standard cosmological paradigm, baryon-cold dark matter interactions are not  universal mechanisms for galaxy formation, and therefore, they cannot be true solutions to the cusp-core problem.
\keywords{cosmology: dark matter --- galaxies: formation --- galaxies: halos --- galaxies: structure --- gravitational lensing}
}
   \maketitle
\section{Introduction}
\label{sect:intro}

In the standard $\Lambda$ cold dark matter cosmology paradigm (called LCDM), the observed large scale structure (LSS) of the universe, which is composed of virialized galaxies and clusters of galaxies, is hierarchically formed from  primordial density fluctuations in the early universe \citep[e.g.,][]{Longair2008}. During the matter dominated epoch, CDM over densities provide deep enough gravitational potentials so that small dark halos can form, merge and evolve more rapidly (compared with only baryons) to make the present LSS. It is believed that, during the whole structure formation process, baryons follow CDM potentials and then reside at the centers of their host dark halos. As a result, each galaxy or cluster of galaxies is surrounded by a dark matter halo. In the framework of LCDM N-body numerical simulations, some analytical and semi-analytical models have successfully predicted  most of the observational properties of LSS. However, challenges arise from small, galactic scales. Numerical simulations predict too many substructures (dwarf galaxies), e.g., several hundreds of satellites are predicted in a galaxy like our Milky Way, while only 23 of them are observed;  too low angular momenta of spiral galaxies; and too high densities (cusps) in the galactic centers \citep{NFW97,Jing2000,Navarro2004,Jing2002,Graham2006}, whereas high resolution observations of the rotation curves for CDM-dominated dwarf and low surface brightness (LSB) galaxies imply that galactic dark matter halos have a density profile with a flat central core \citep{deBlok2001b,van2001,Swaters2003,Weldrake2003,Donato2004,Gentile2005,Simon2005,Gentile2007}. The latter is known as the cusp-core problem, which we will investigate in detail in this paper. 

Many efforts have been devoted to resolve the challenges mentioned above. The less traditional models include warm dark matter and self-interacting dark matter. In the LCDM model, some researchers simply deny the validity of the observations, others introduce baryon-cold dark matter interactions (BCDMIs) such as dynamical friction of substructures \citep{El-Zant2001,Tonini2006,Romano2008}, stellar bar-CDM interaction \citep{Weinberg2002,Holley2005}, and baryon energy feedback \citep{Mashchenko2006, Peirani2008}. In this paper, we focus on the validity of BCDMI solutions to the cusp-core problem.

First of all, we point out that any reasonable solutions to the cusp-core problem should be verified by the observations of galaxies not only on small scales like dwarf and LSBs, but rather on all galactic scales, especially on scales like giant ellipticals. Why should we include the giant ellipticals when testing structure formation models (e.g., the solutions to the cusp-core problem)? According to the hierarchical structure formation theory in the frame work of LCDM, small CDM halos form first, then merge to make increasingly large halos. While N-body numerical simulations of pure CDM with higher and higher force and mass resolution still favor cuspy halo profiles, the BCDMIs introduced in the baryon plus CDM regime can transfer the early formed cusp into a central flat core in later stages for each dark halo. It is well known that the ``universal'' density profile known as NFW \citep[Navarro-Frenk-White;][]{NFW97} is valid for halos ranging from dwarf galaxies to those as large as clusters of galaxies. The only difference between pure CDM and baryon plus CDM regimes is that the latter considers the effects of baryons. Their formation processes should be similar (bottom-up or hierarchical) in the spirit of LCDM. Therefore, the properties of the galaxies (e.g., the slope in the central region) predicted by BCDMIs should be valid and meaningful at all mass scales. Needless to say, for any theoretical models describing structure formation, it is meaningless or impossible to customize some physical processes only to explain the observations of dwarfs and LSBs, and exclude ellipticals as exceptions. In other words, no matter what galaxy types the initial conditions presumed, the subsequential and final (present) CDM halos predicted by simulations should include all galaxy types and have the same observational properties. This is why we should include giant ellipticals when testing  solutions to the cusp-core problem. Usually, when we talk about the cusp-core problem, we mean the contradiction between the ``cusp'' predicted by simulations of pure CDM and the ``core''implied by rotation curves of dwarfs and LSBs, as is widely known. However, we forget the fact that giant elliptical galaxies are also important members of the galactic family, but they prefer a cuspy (or small core-size) density profile in their central region according to observations (e.g., x-rays and the statistics of strong gravitational lensing). In deed, LSB galaxies and giant ellipticals are quite different galaxies; the former are CDM-dominated while the latter have baryon-dominated centers. However, their distinguishing differences are obtained from all kinds of observational data rather than  theoretically predicted from structure formation theories. It is theories that should successfully explain such observational differences, not the other way around. The density profiles of giant ellipticals with baryon-dominated centers (a cusp or small core)  should be predicted by the same model, if reasonable, which predicts the density profile of LSB galaxies with CDM-dominated centers (a large flat core).

It is interesting to note that some solutions to the cusp-core problem based on analytical models are problematic in a similar way to simulations. One such model assumes that cosmological halos form from the collapse and virialization of ``top-hat'' density perturbations and are spherical, isotropic and isothermal. This predicts a unique, nonsingular, truncated isothermal sphere (NTIS) and provides a simple physical clue about the existence of soft cores in halos of cosmological origin \citep{Shapiro1999,Iliev2001}. It is claimed that this NTIS model is in good agreement with observations of the internal structure of dark matter-dominated halos on scales ranging from dwarf galaxies to X-ray clusters. Unfortunately, the NTIS model is ruled out by the statistics of strong gravitational lensing \citep{Chen2005}.

In summary, any self-consistent, reasonable solutions to the cusp-core problem, either numerical or analytical, should be able to successfully explain the cuspy density profiles of giant ellipticals simultaneously, otherwise, no matter how reasonable it looks, such a solution cannot be true.  

While giant ellipticals should be included, now the question is: to what extent is the core size of the giant ellipticals allowed by observations? In particular, is it acceptable to extrapolate the core size-halo mass relation constrained by the rotation curves of dwarfs and LSB galaxies to giant ellipticals? Like \citet{Chen2008}, we investigate these problems by ``reasonably'' extrapolating the core-mass relation constrained from LSB galaxies to strong lensing galaxies (usually ellipticals), and comparing the predicted lensing frequency with observations. Since it was explicitly pointed out in \citet{Mashchenko2006} that the CDM density profile of the evolved halo is in remarkable agreement with the Burkert profile, we fit the observational data of rotation curves with the Burkert density profile in section \ref{sect:RCs}, the corresponding lensing probabilities are presented in section \ref{sect:lens}, and we give our discussion and conclusions in section \ref{sec:discussion}. 

\section{Density profile constrained from observations of rotation curves}
\label{sect:RCs}

We use high-resolution and high-quality hybrid $\textrm{H}_{\alpha}$/H I rotation curves of a sample of 26 LSB galaxies, analyzed in \citet{deBlok2001a}, to fit the parameters in the Burkert density profile \citep{Burkert1995}. Some curves of this sample were taken from the large sample of 50 LSB galaxies presented in \citet{McGaugh2001}. While these curves were well fitted by the cored isothermal sphere  model \citep{Begeman1991,deBlok2002,deNaray2008}, we will show bellow that they are also well fitted by the Burkert density profile \citep[see also][]{Salucci2000,Gentile2004,Salucci2007}. As a typical model of BCDMIs, turbulence driven by stellar feedback during galaxy formation is a possible solution to the cusp-core problem \citep{Mashchenko2006}. Numerical N-body simulations show us that random bulk motions of gas in small primordial galaxies result in a flattening of the central CDM cusp. Phase space arguments imply that the core should persist through subequent mergers \citep{Dehnen2005,Kazantzidis2006}. Consequently, in the present universe both small and large galaxies would have flat CDM core density profiles, which can be well fitted with the Burkert model \citep{Burkert1995,Mashchenko2006},
\begin{equation}
\rho(r)=\frac{\rho_{0}r_{0}^{3}}{\left( r+r_{0}\right)(r^{2}+r_{0}^{2})},
\end{equation} 
where $\rho_0$ is the central density near $r=0$ and $r_0$ is the core radius, the two free parameters to be fitted. The mass within radius $r$ is \citep{Salucci2007},
\begin{equation}
 M(r)=2\pi\rho_0r_0^3\left\{\ln\left(1+\frac{r}{r_0}\right)+
\ln\left[1+\left(\frac{r}{r_0}\right)^2\right]-\arctan\left(\frac{r}{r_0}\right)\right\}. 
\label{Mr}
\end{equation}

\begin{figure}[!hb]
\centering
\includegraphics[width=0.2\textwidth]{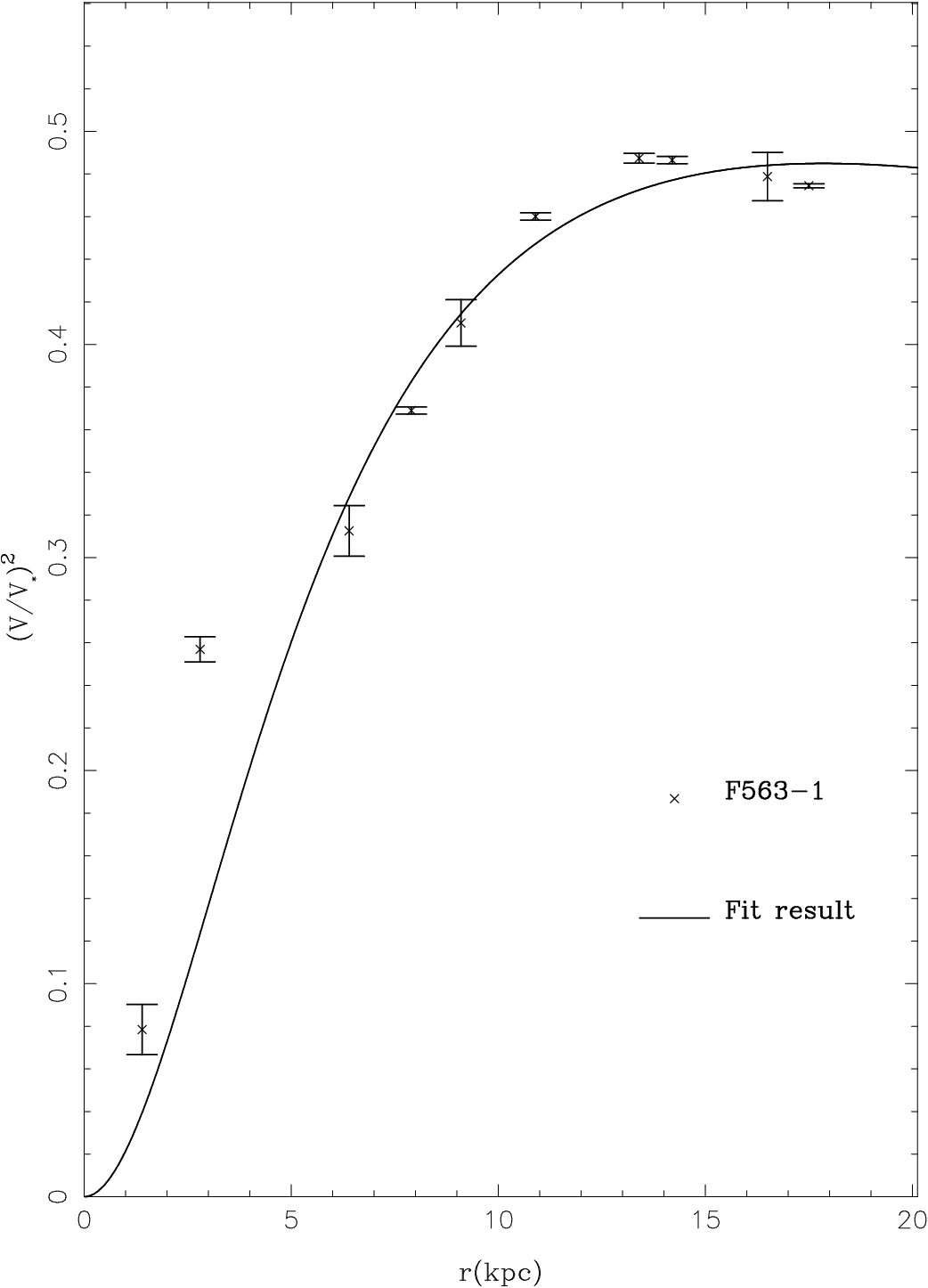}
\includegraphics[width=0.2\textwidth]{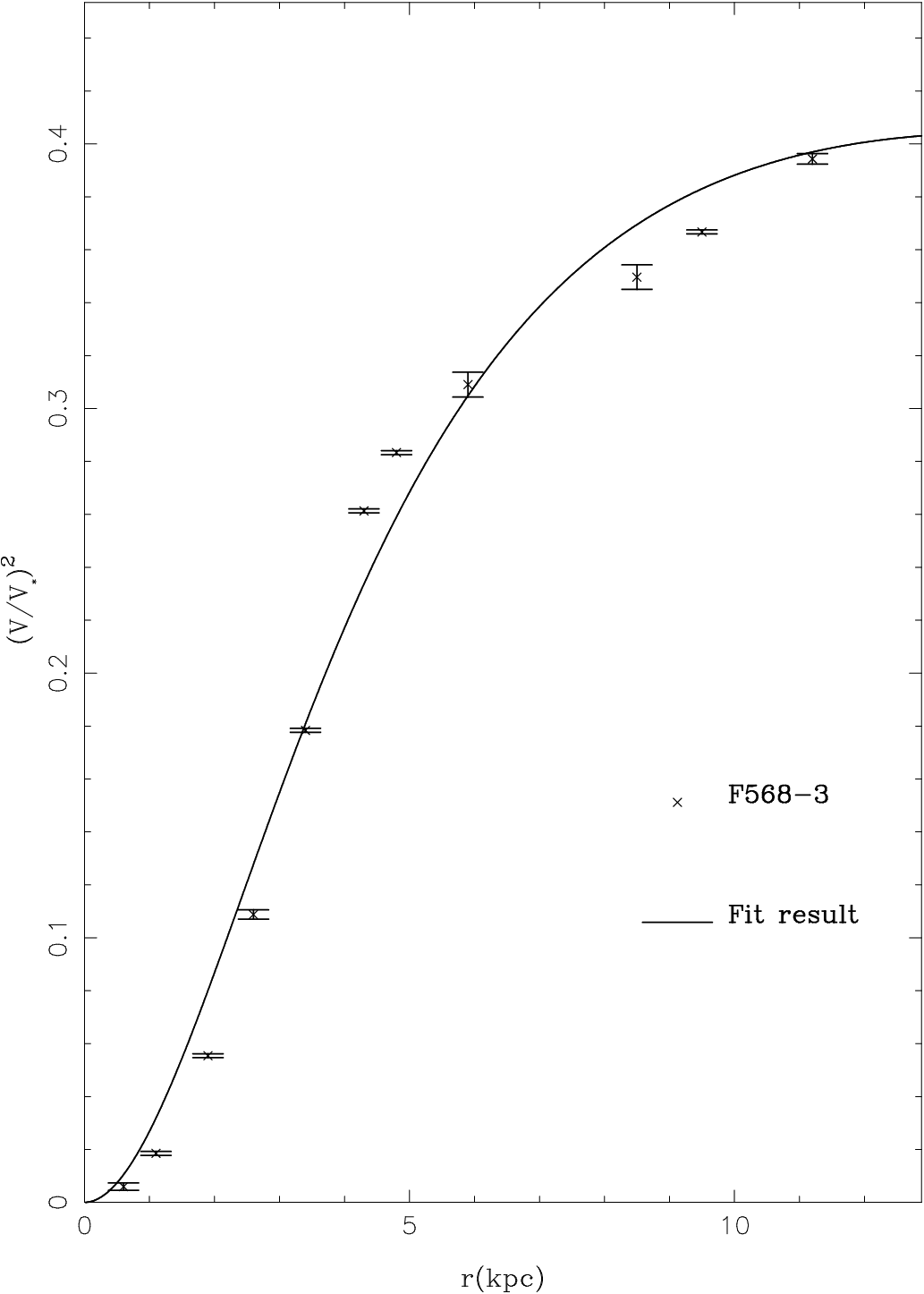}
\includegraphics[width=0.2\textwidth]{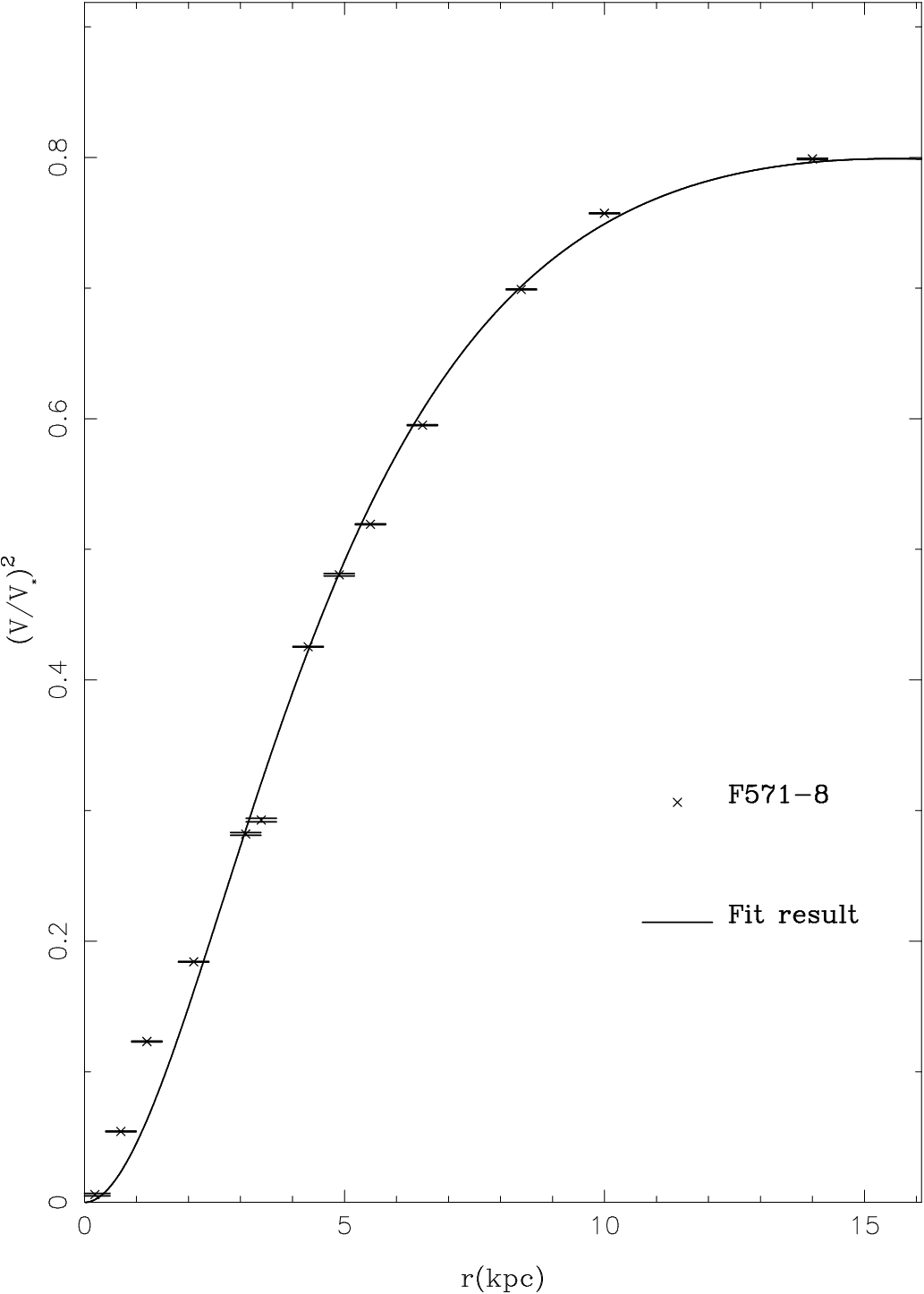}
\includegraphics[width=0.2\textwidth]{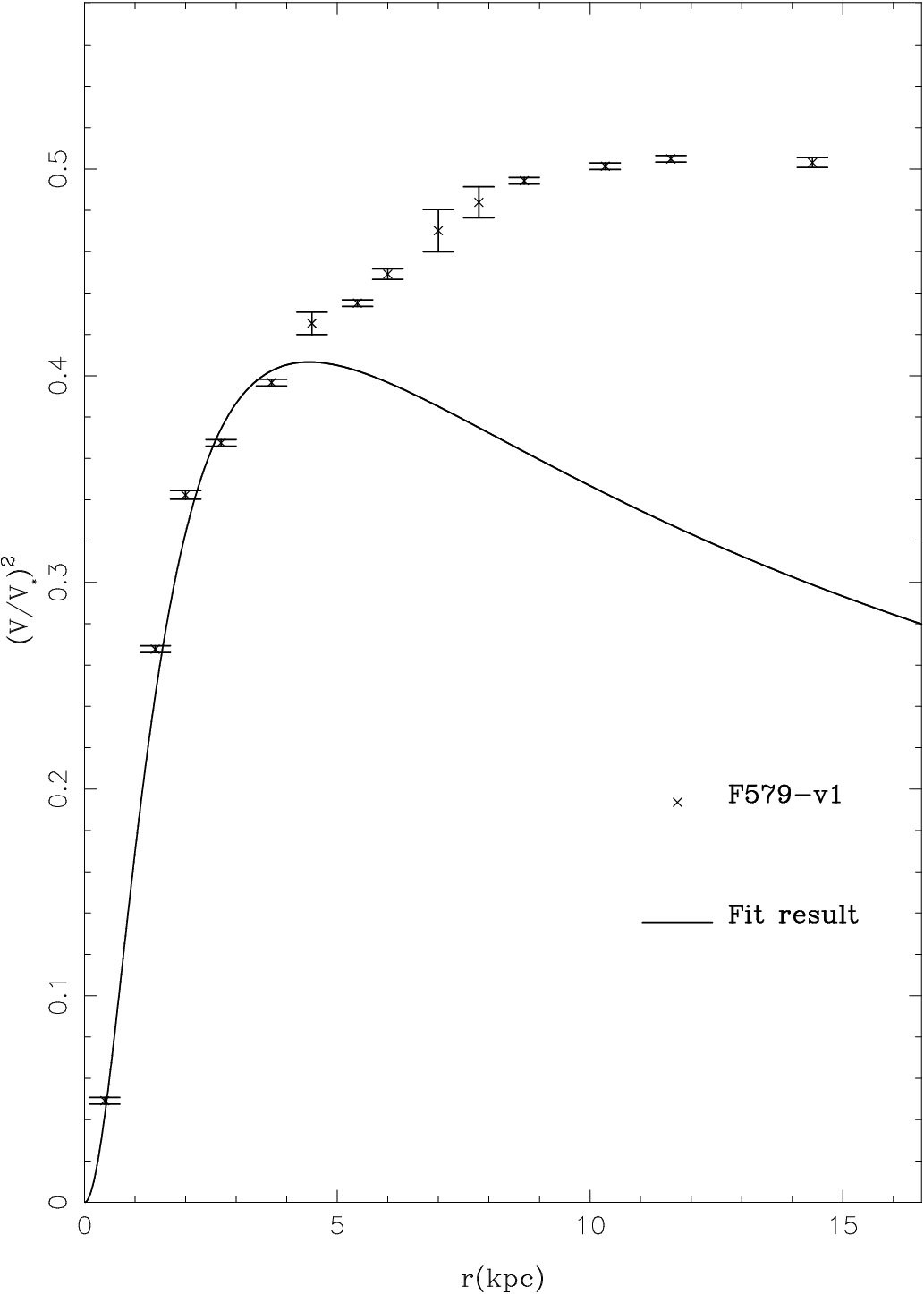} \\
\includegraphics[width=0.2\textwidth]{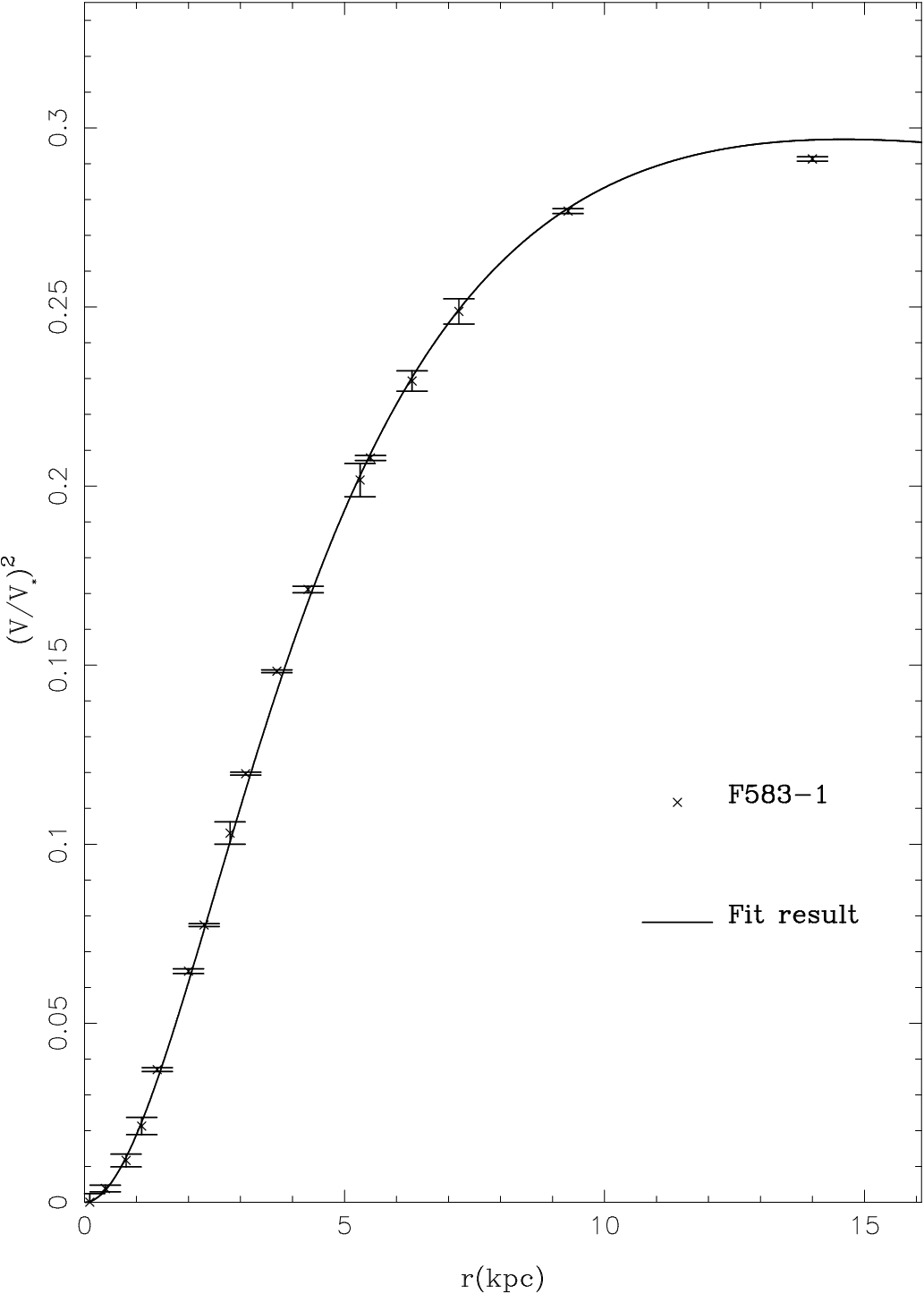} 
\includegraphics[width=0.2\textwidth]{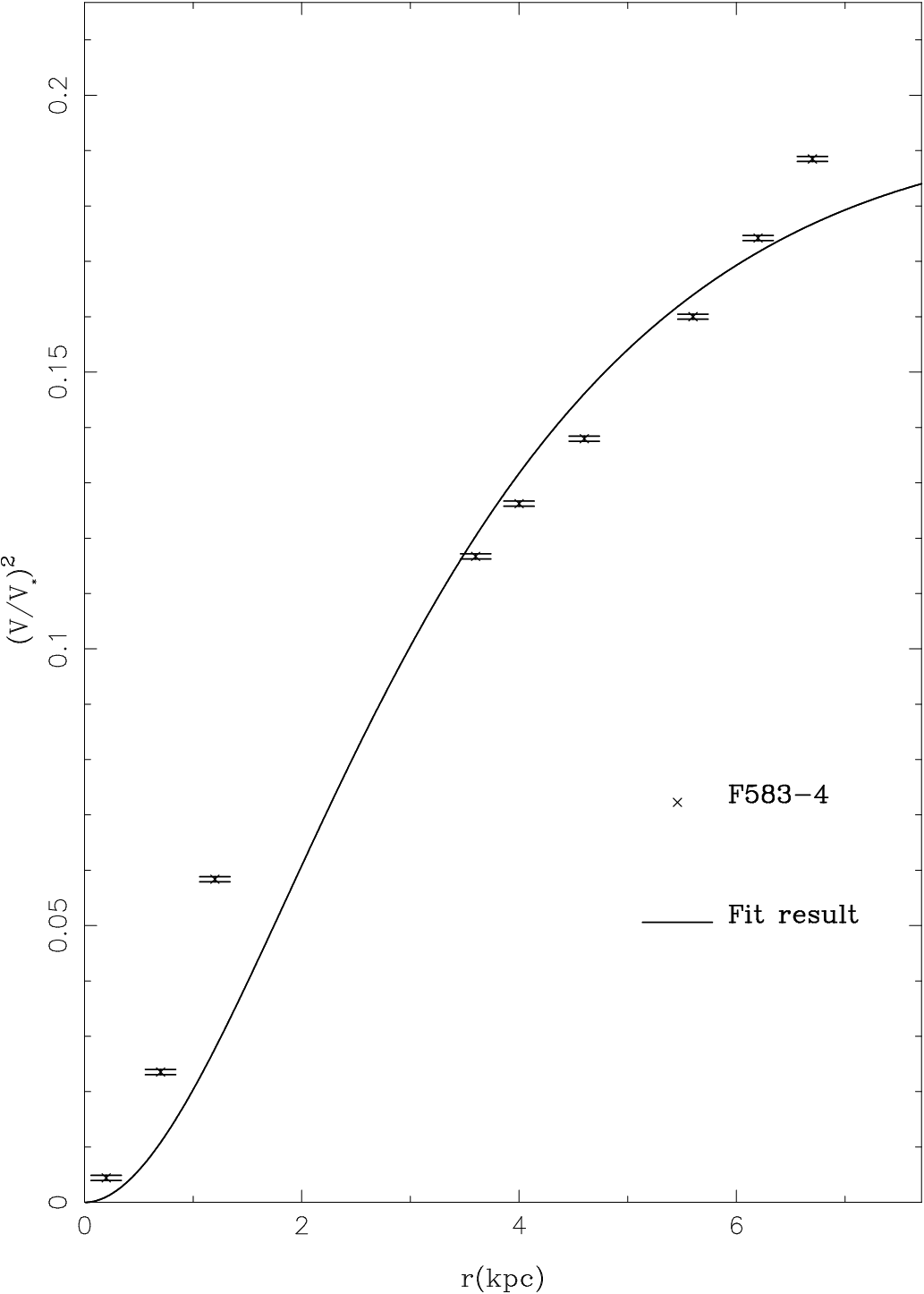} 
\includegraphics[width=0.2\textwidth]{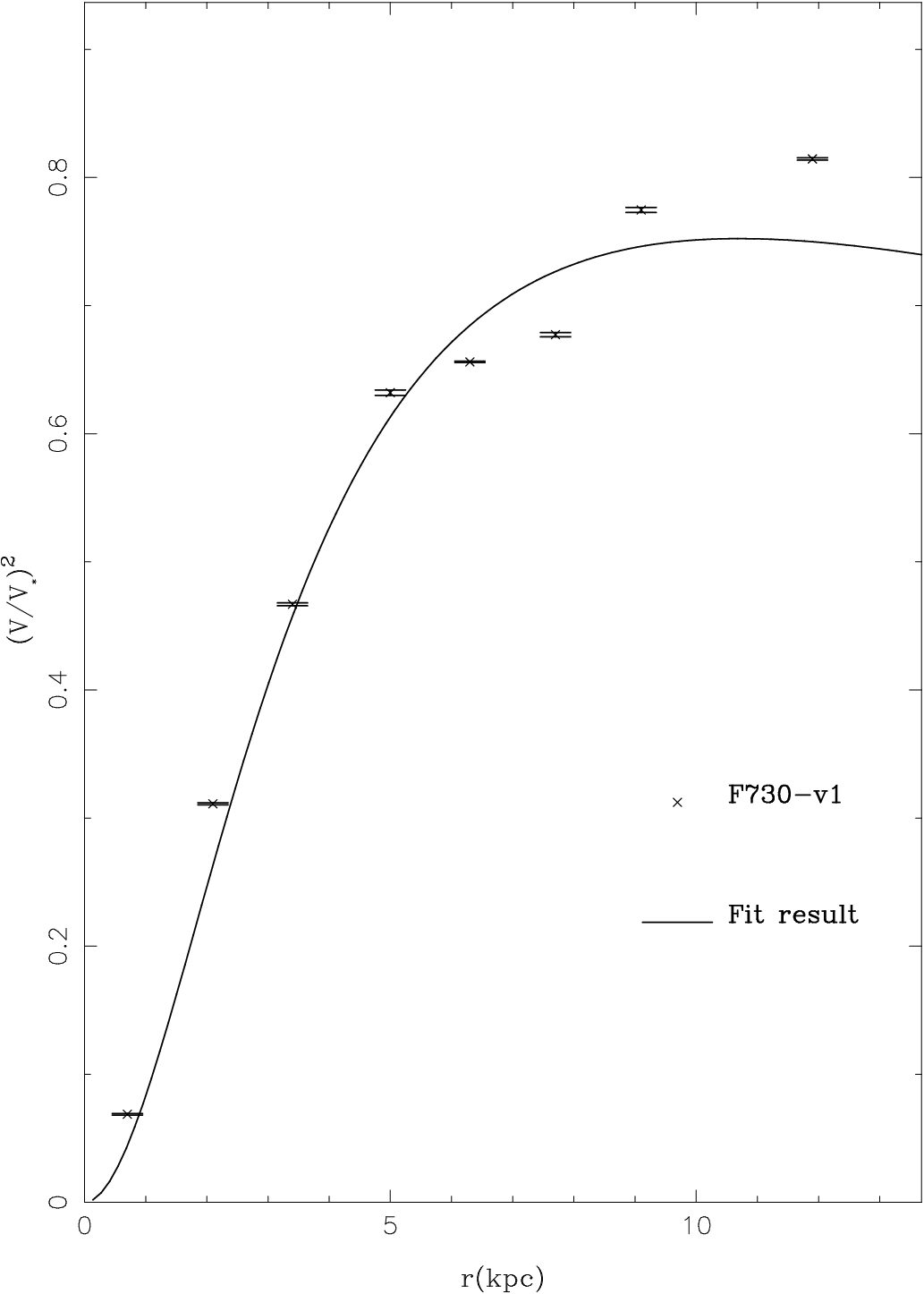}
\includegraphics[width=0.2\textwidth]{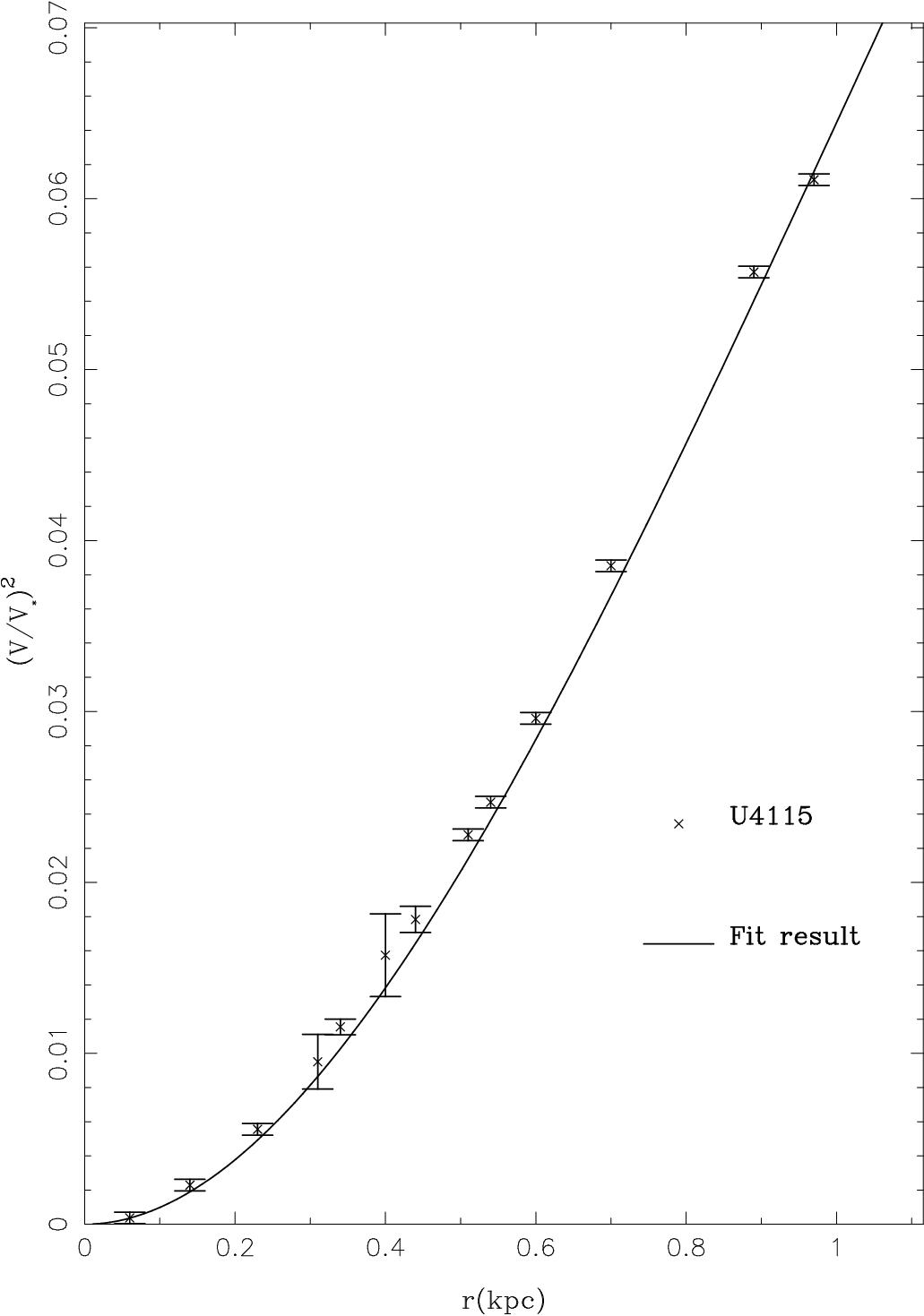} \\
\includegraphics[width=0.2\textwidth]{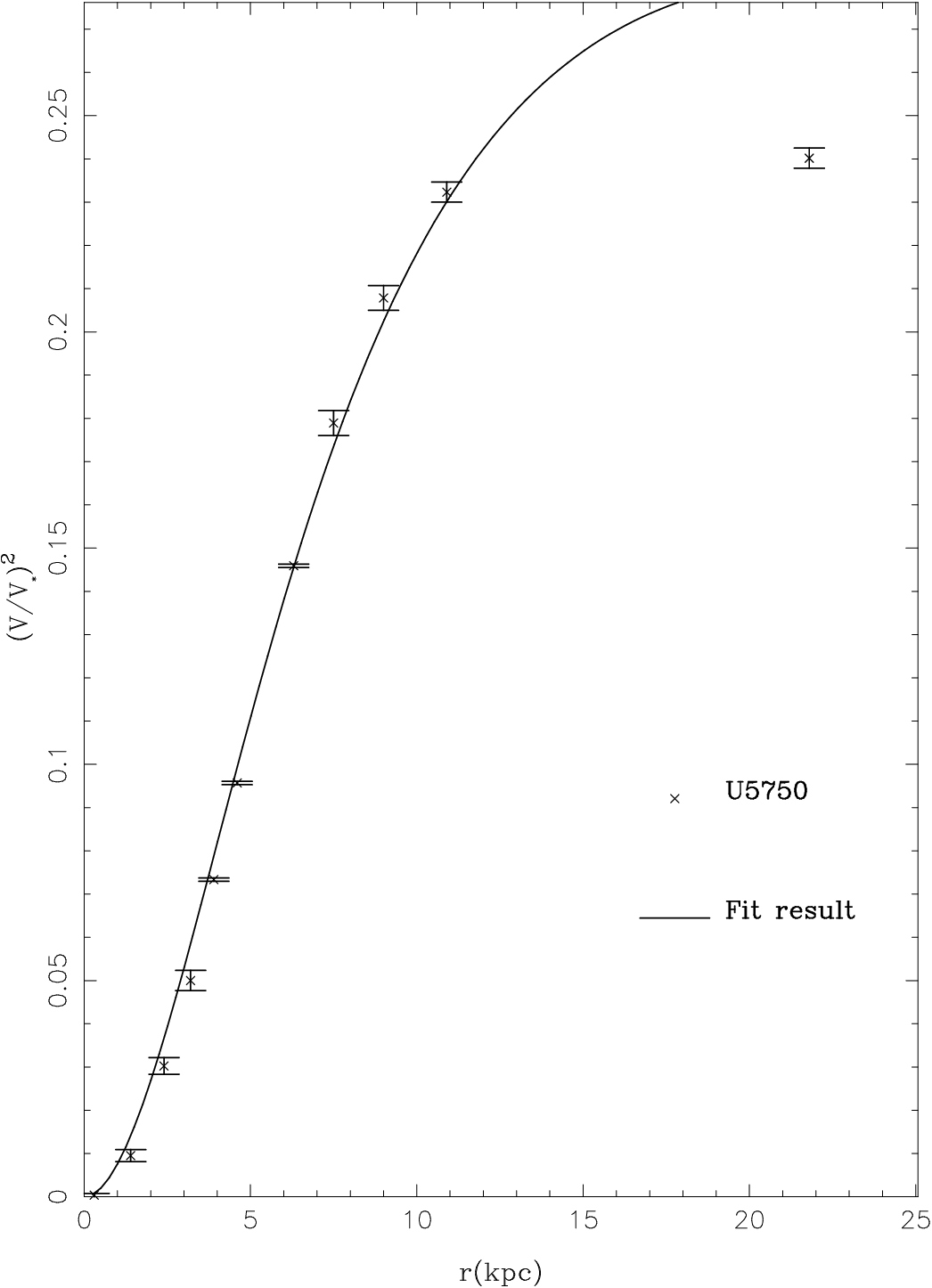}
\includegraphics[width=0.2\textwidth]{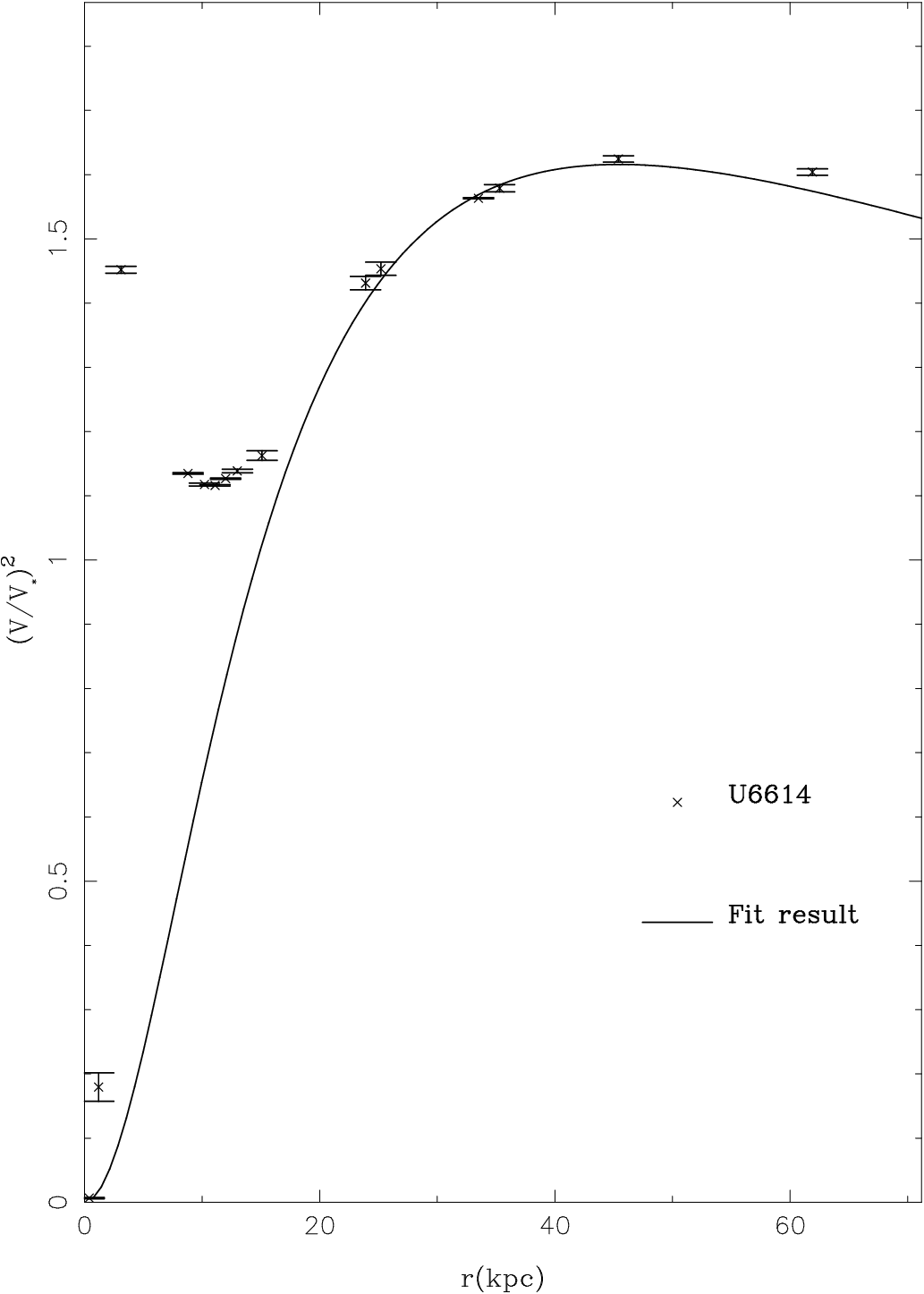}
\includegraphics[width=0.2\textwidth]{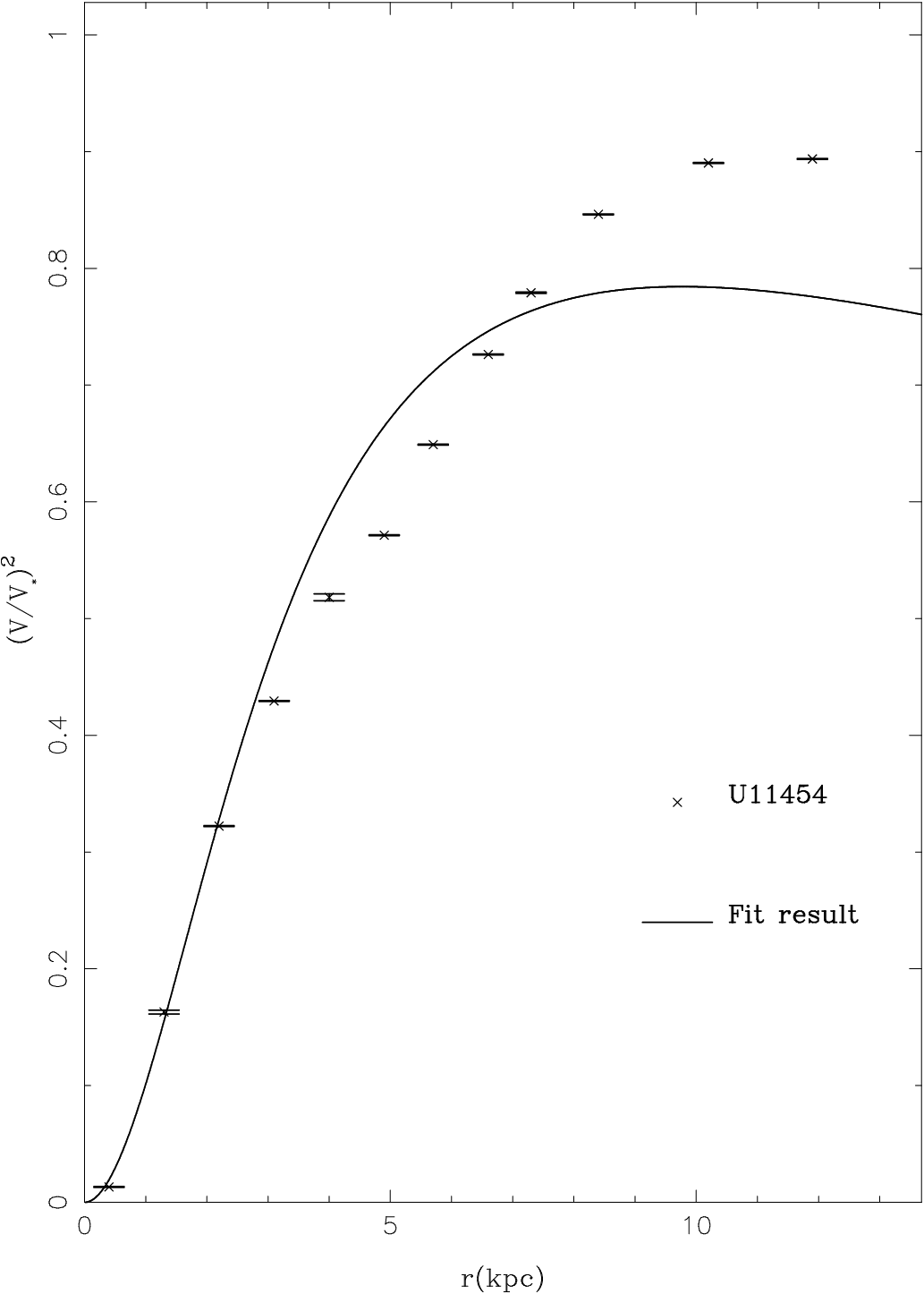}
\includegraphics[width=0.2\textwidth]{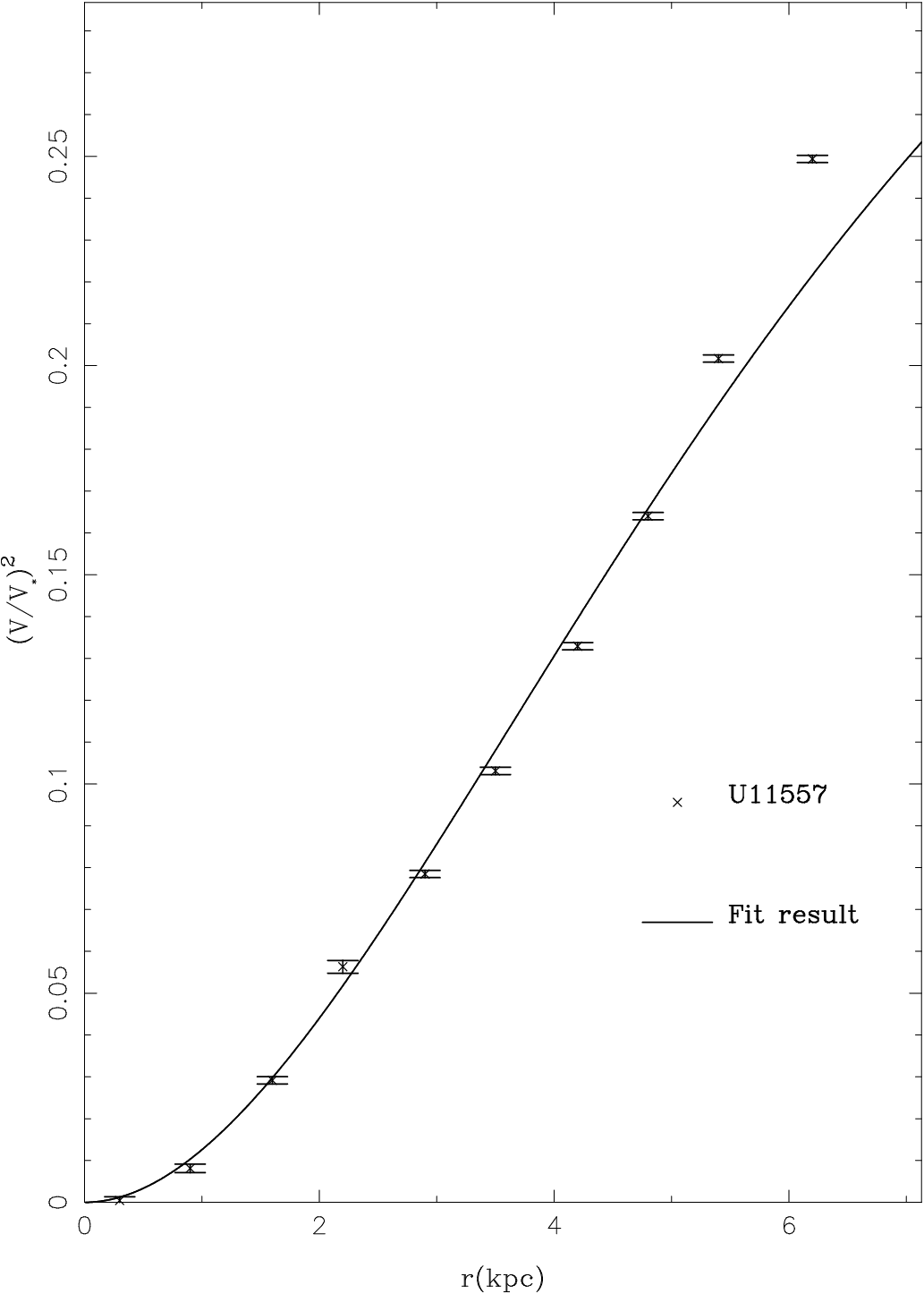} \\
\includegraphics[width=0.2\textwidth]{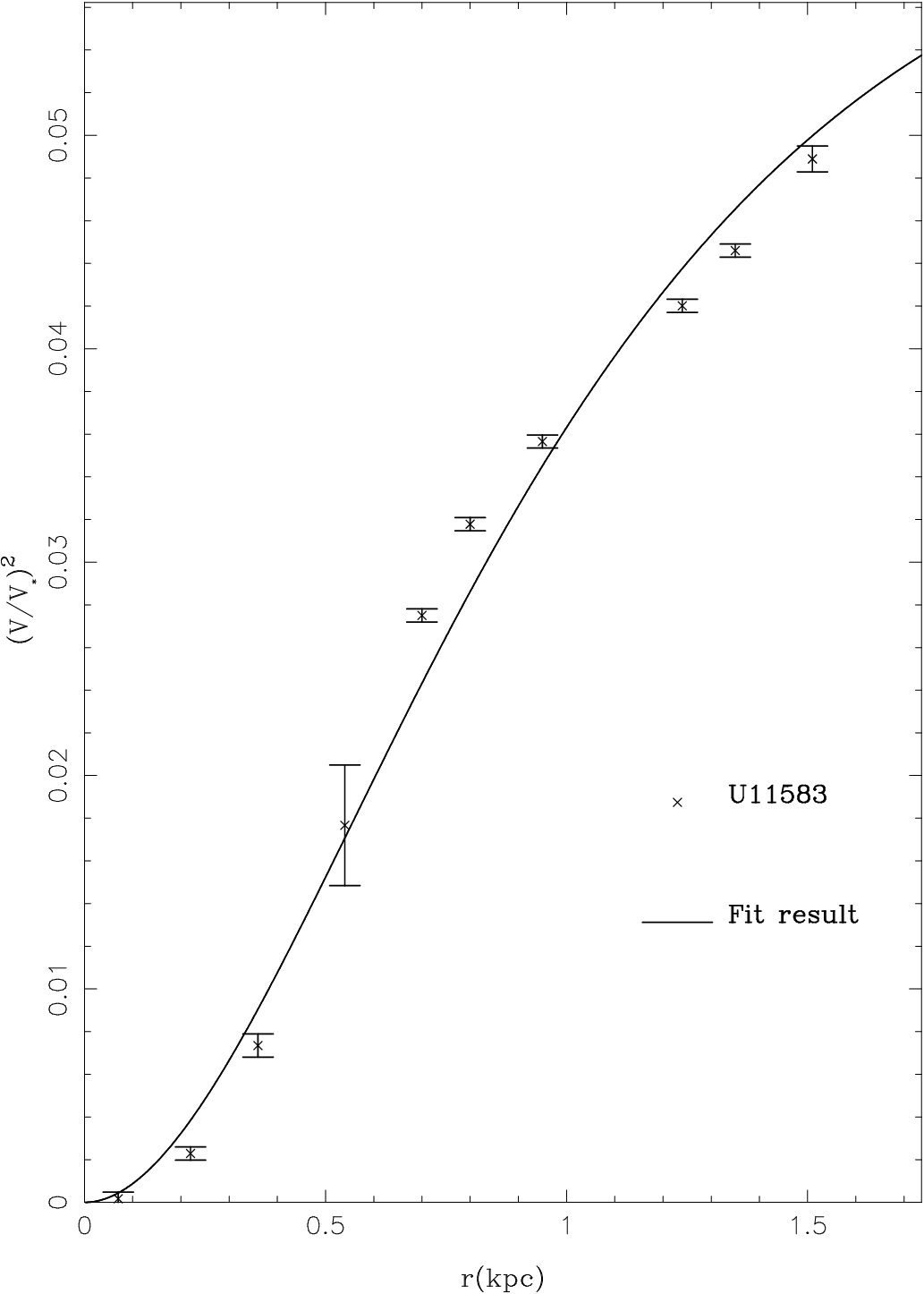}
\includegraphics[width=0.2\textwidth]{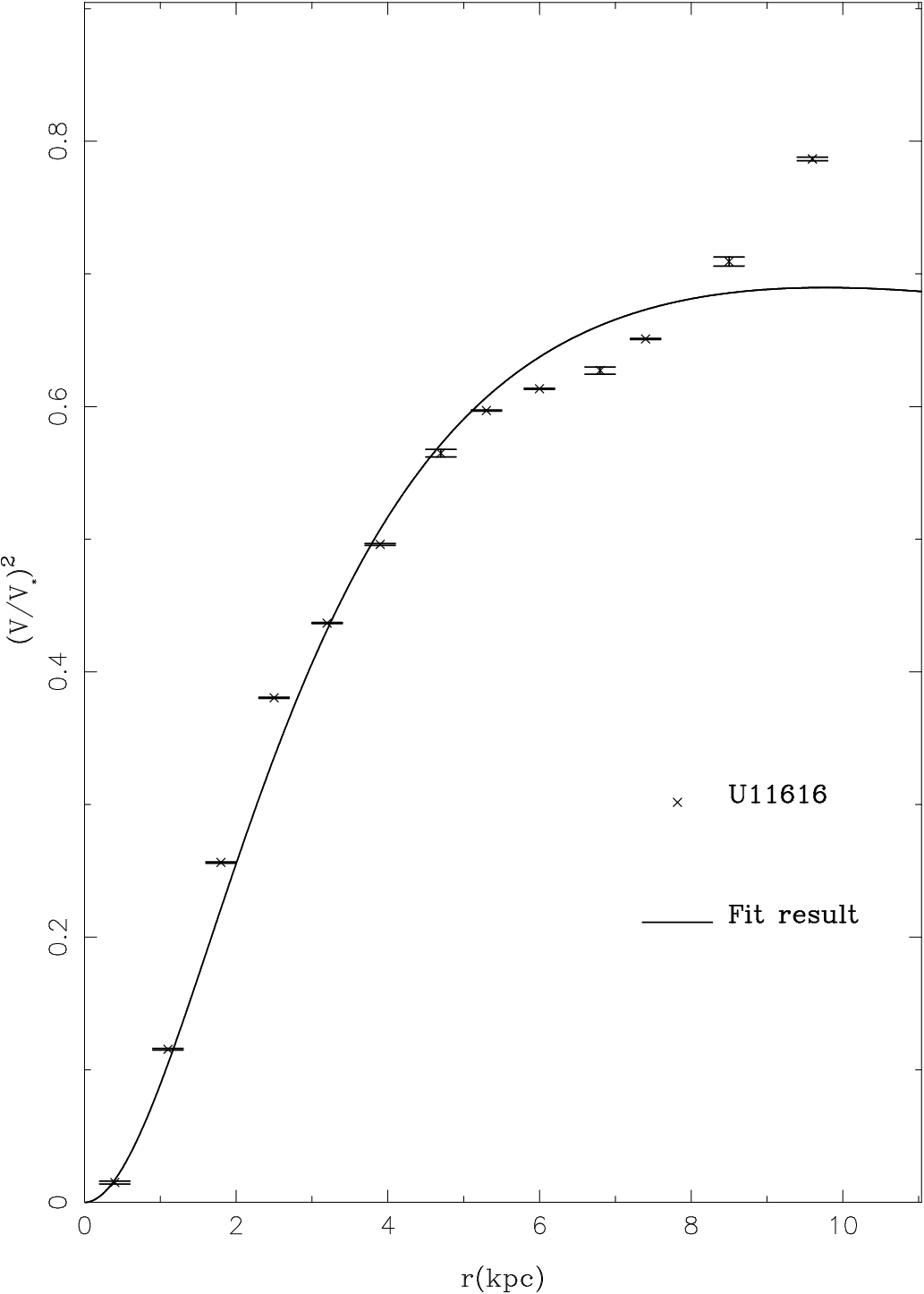}
\includegraphics[width=0.2\textwidth]{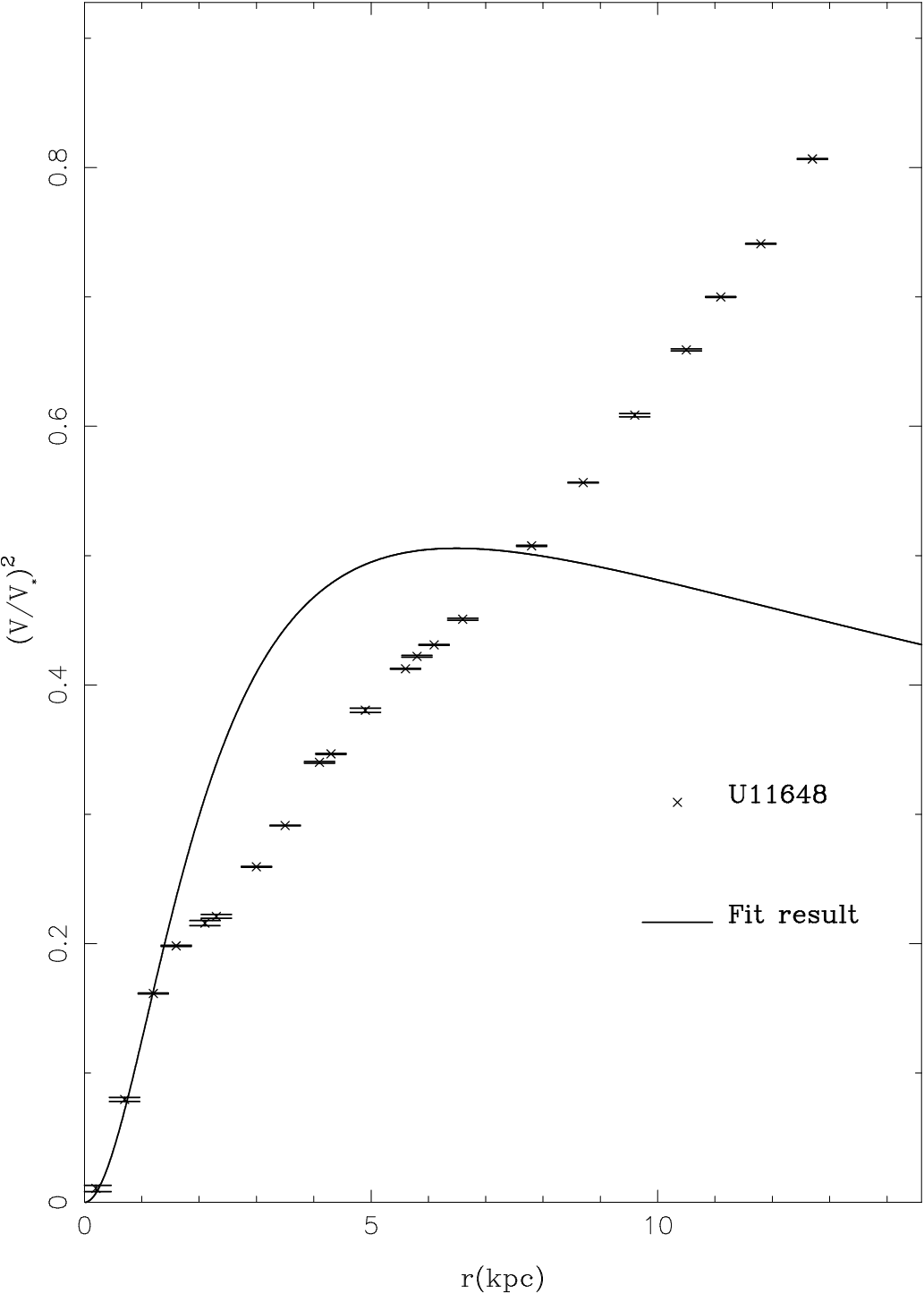}
\includegraphics[width=0.2\textwidth]{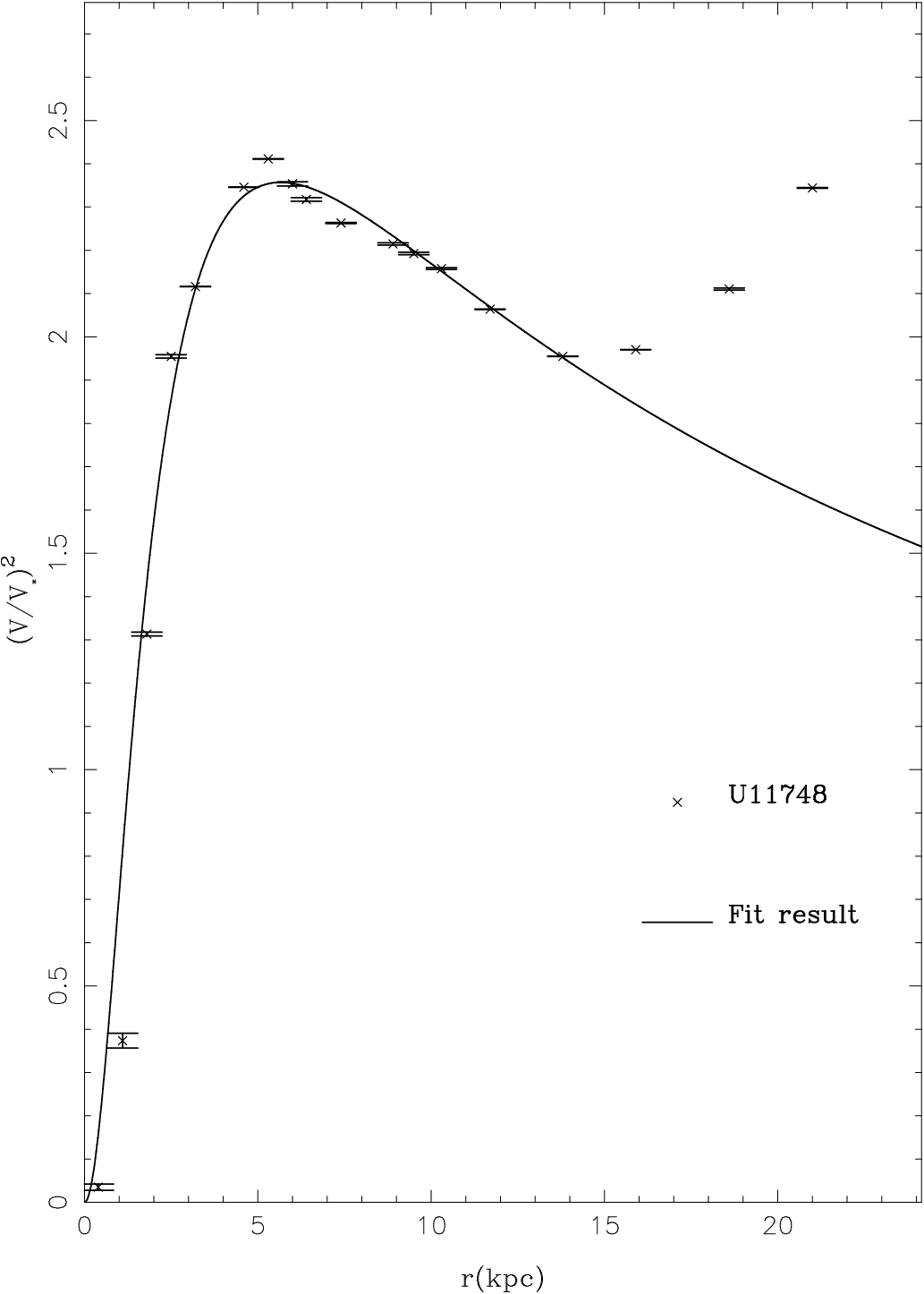} 
\caption{Comparison of the fitted $V^2(r)$ derived from the Burkert profile (solid lines) and the observed rotation curves (crosses).}
\label{Vr}
\end{figure}
\addtocounter{figure}{-1}
\begin{figure}[!ht]
\includegraphics[width=0.2\textwidth]{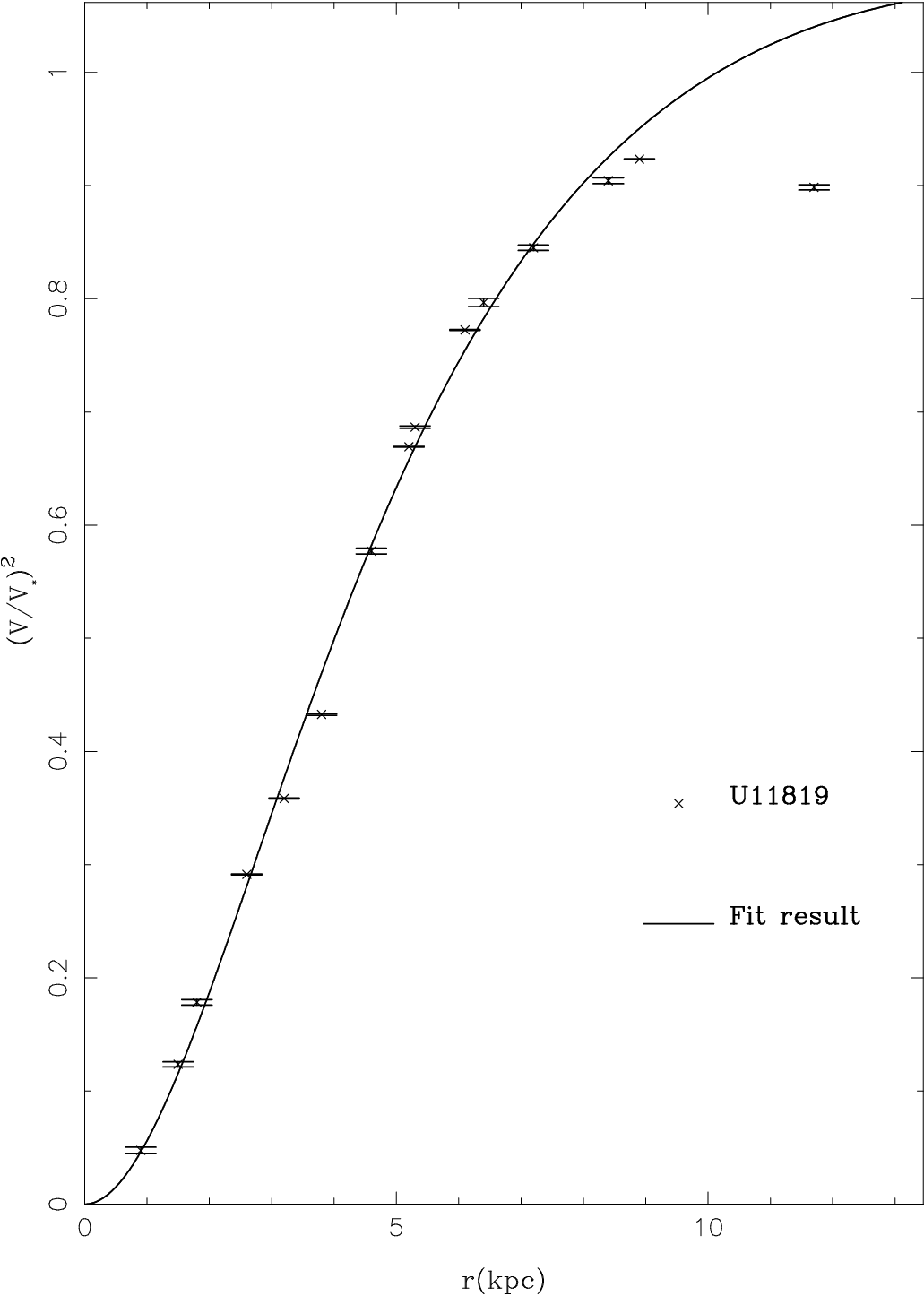}
\includegraphics[width=0.2\textwidth]{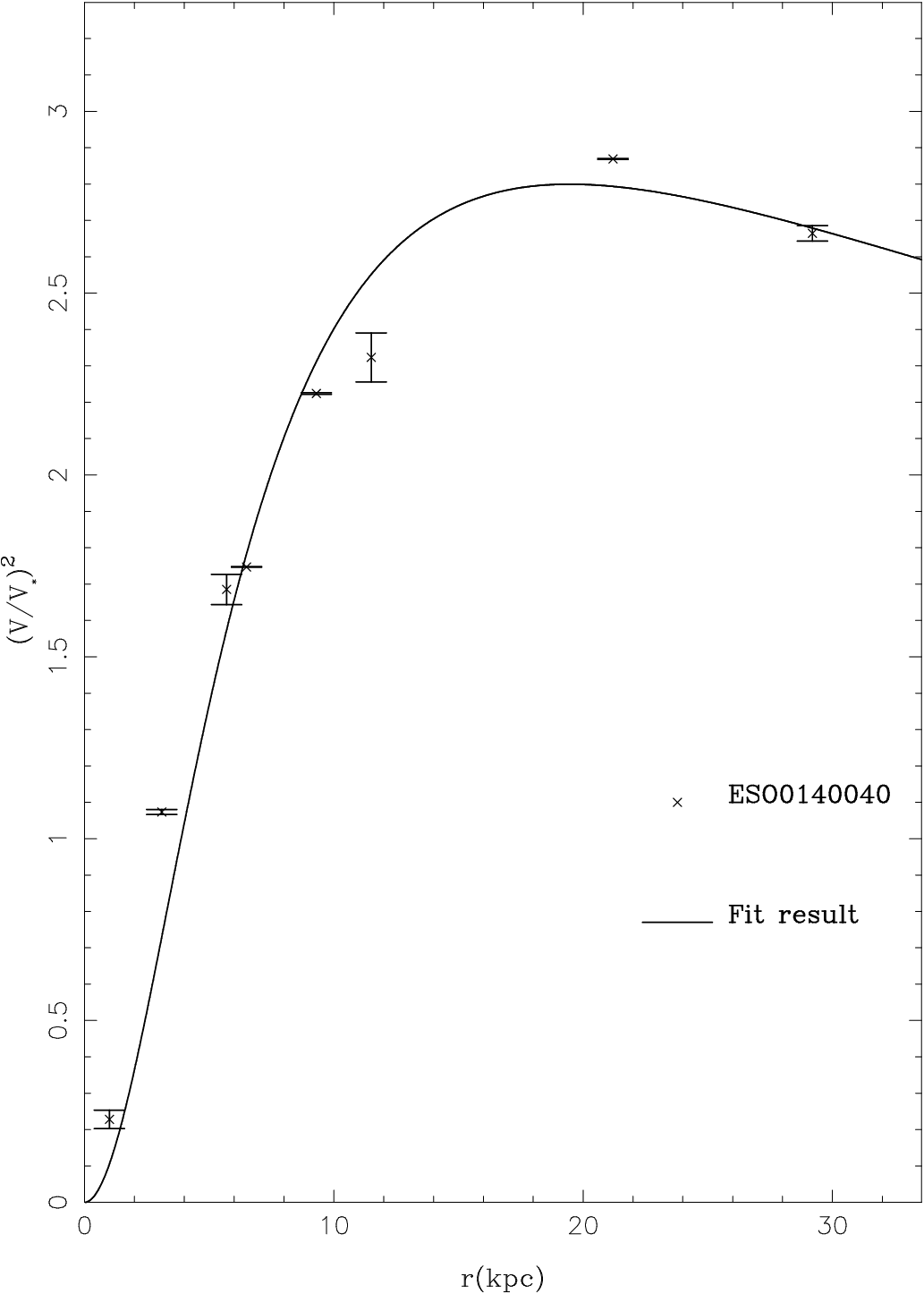} 
\includegraphics[width=0.2\textwidth]{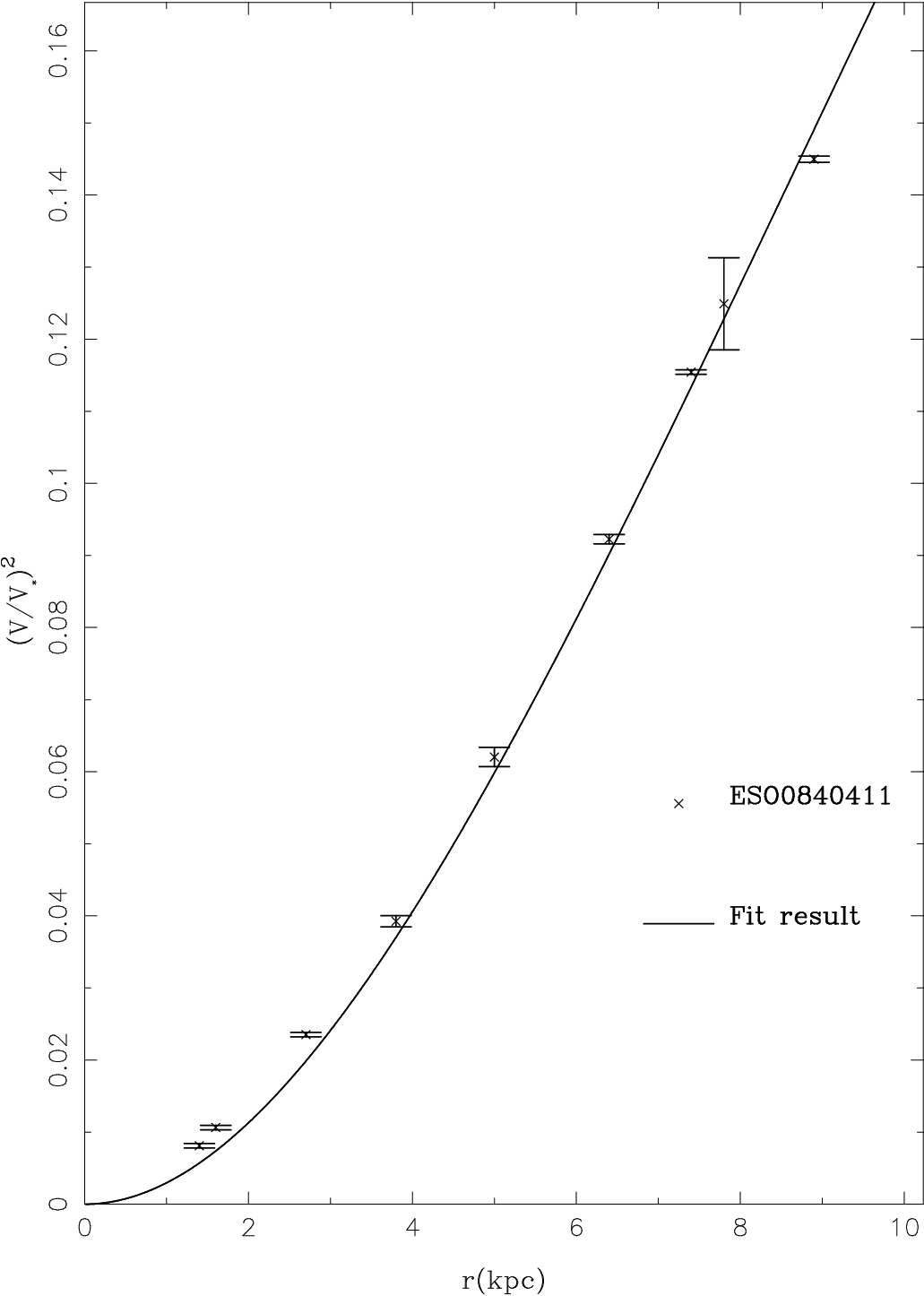}
\includegraphics[width=0.2\textwidth]{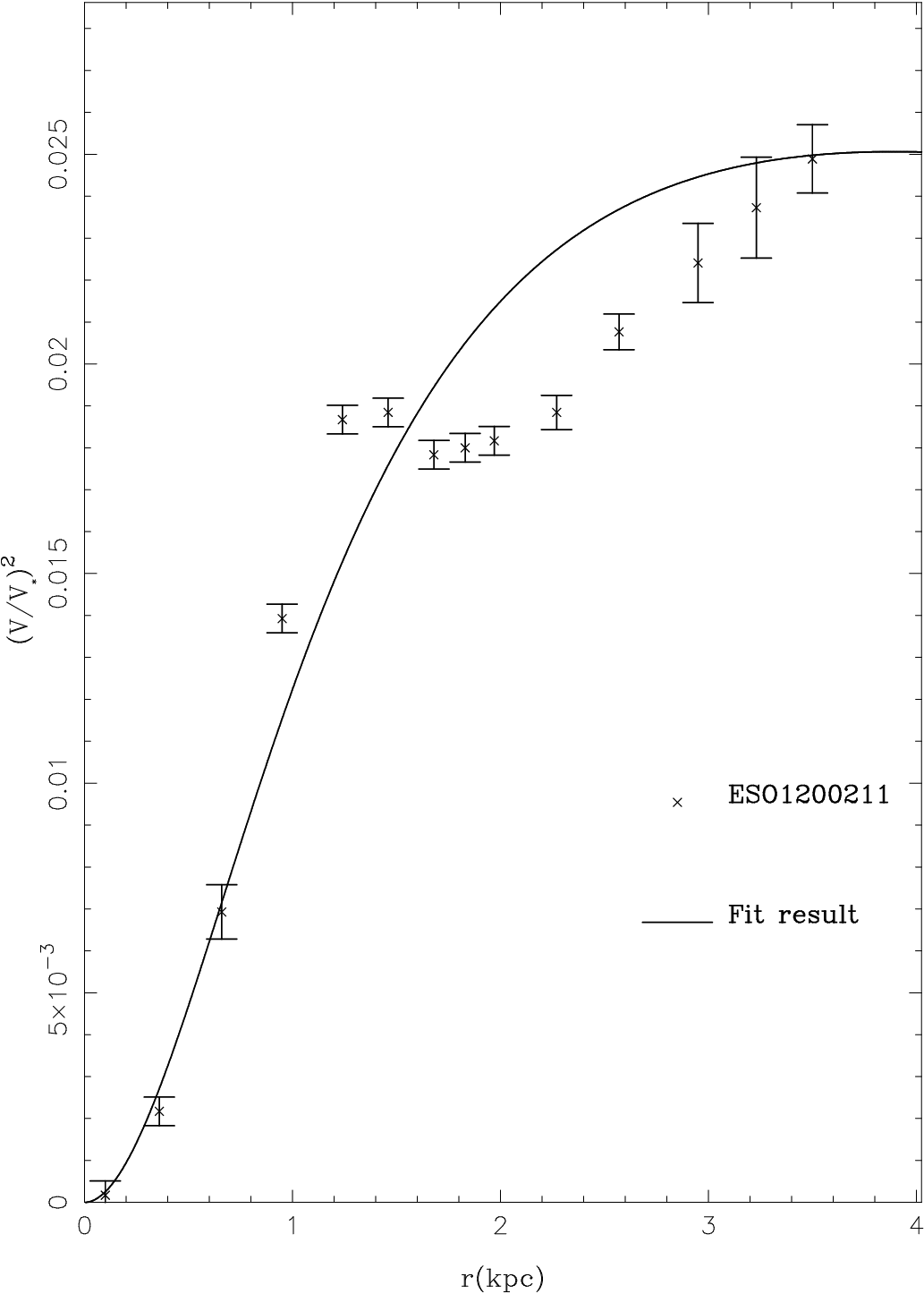} \\
\includegraphics[width=0.2\textwidth]{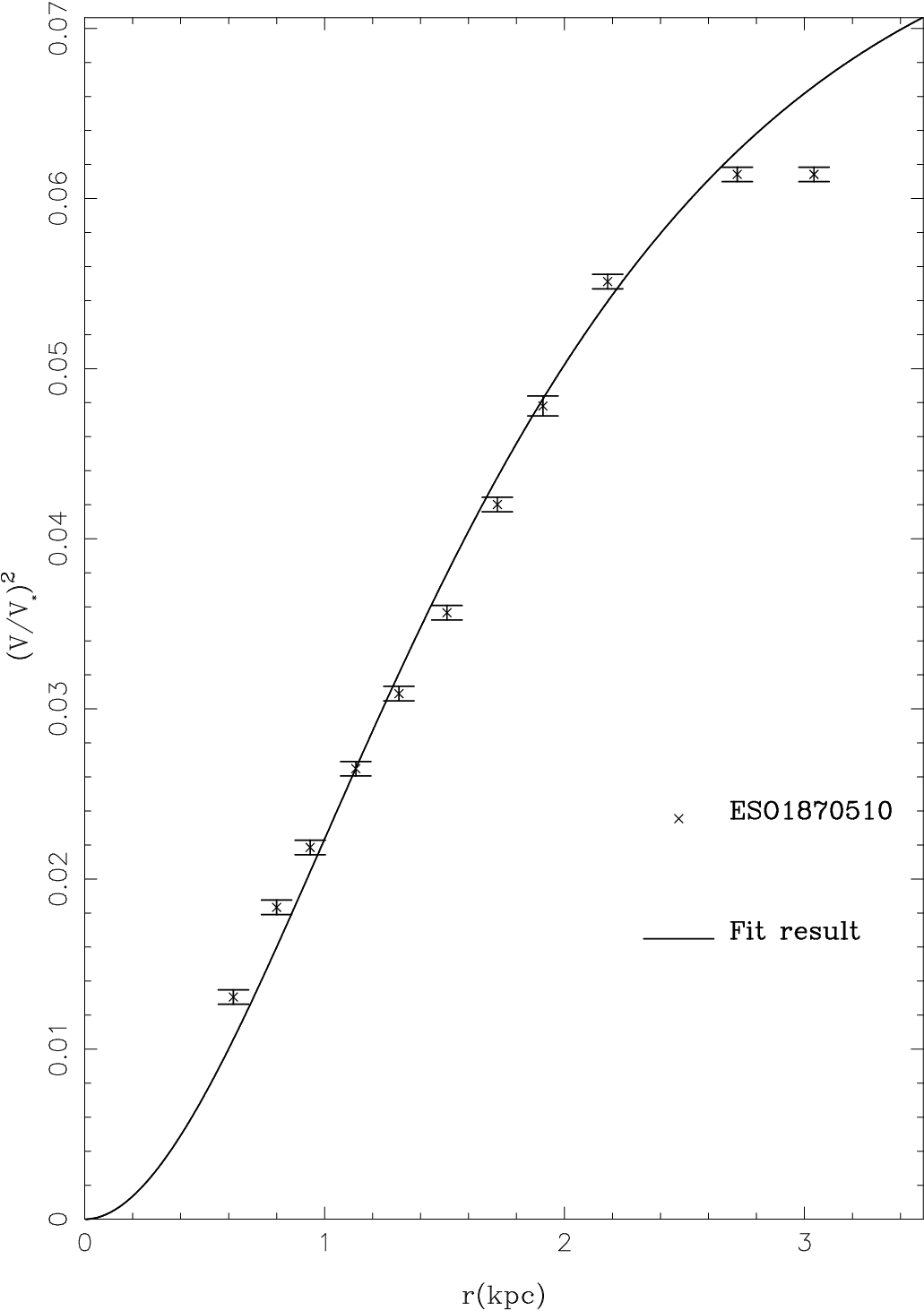}
\includegraphics[width=0.2\textwidth]{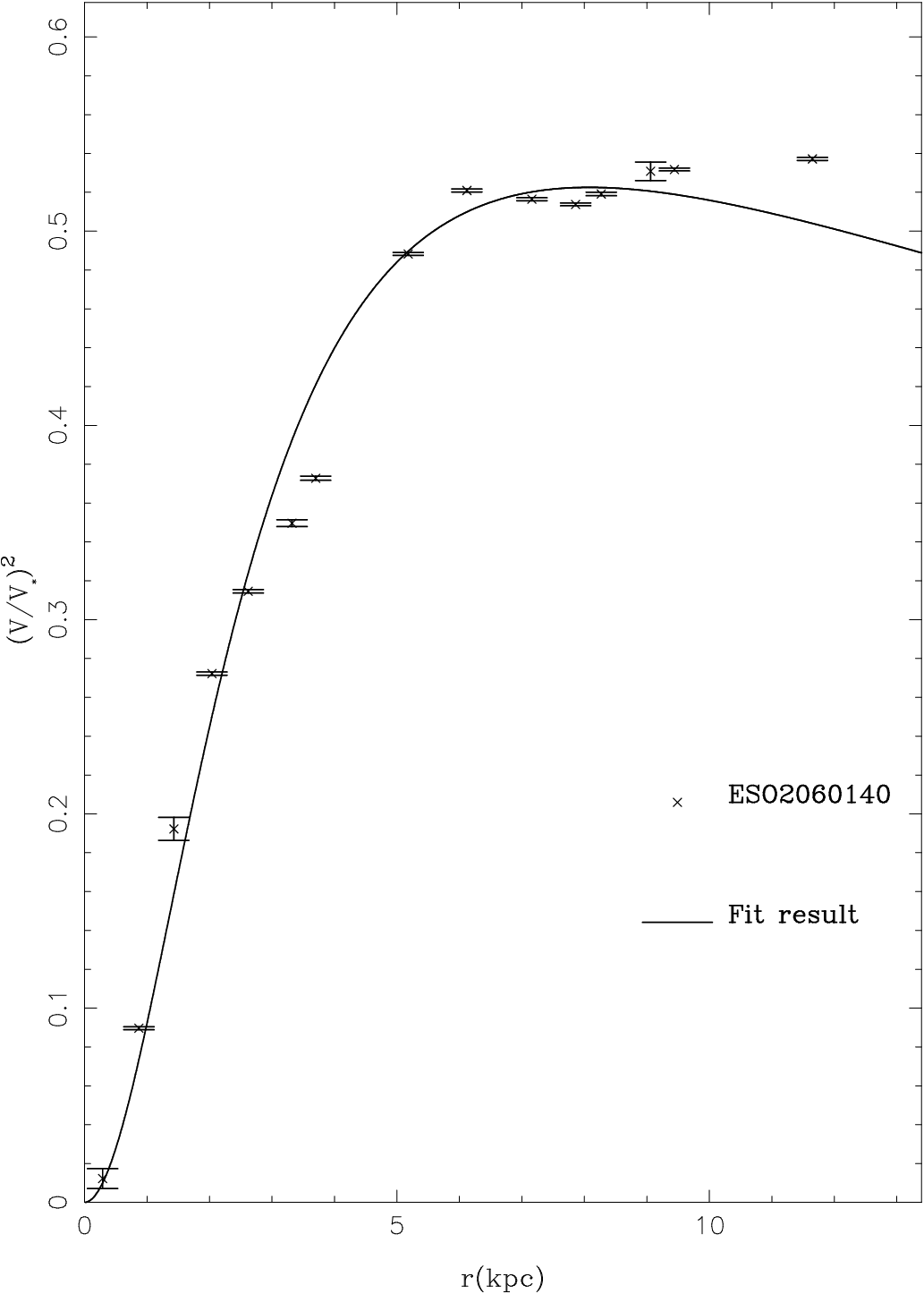}
\includegraphics[width=0.2\textwidth]{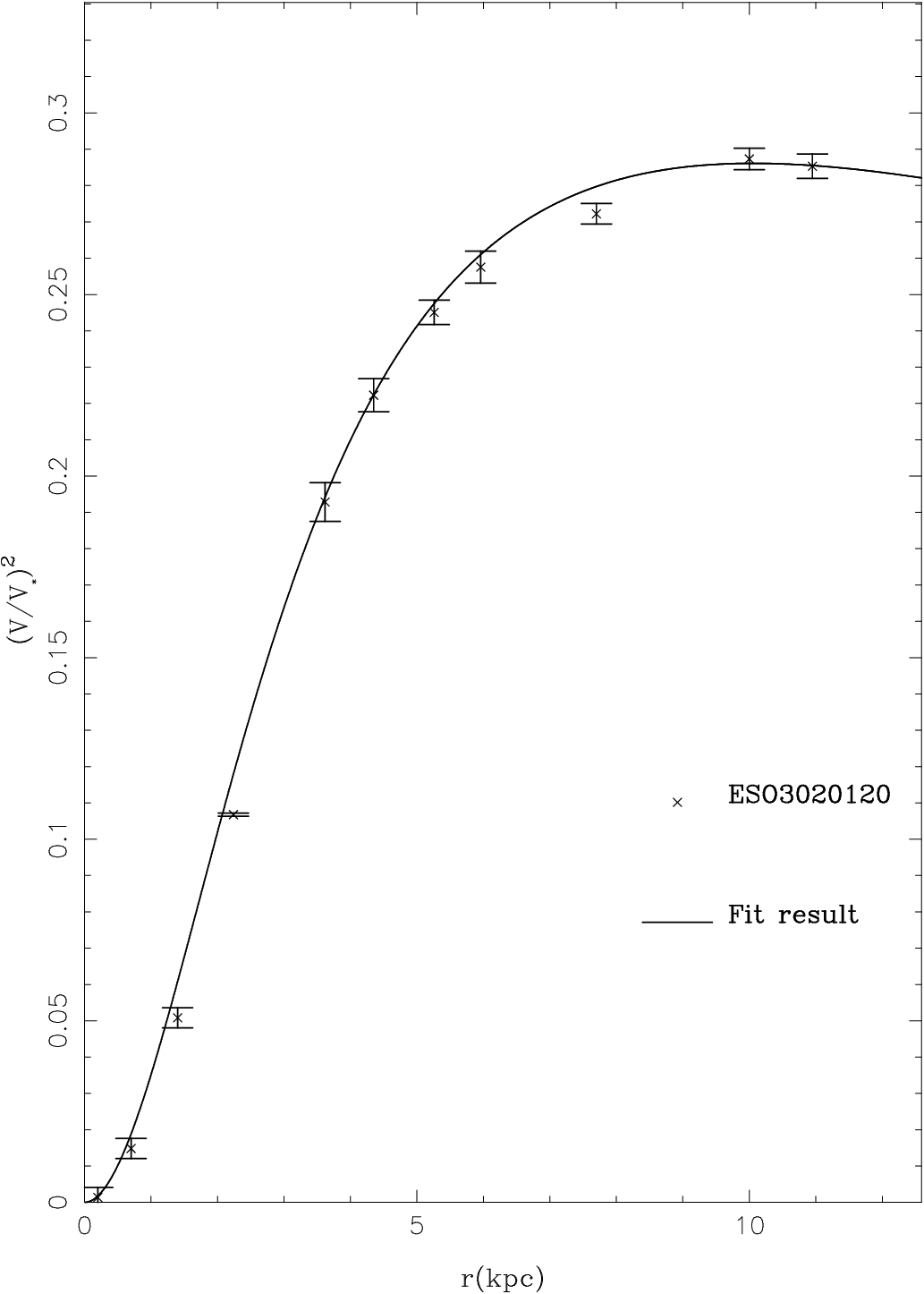}
\includegraphics[width=0.2\textwidth]{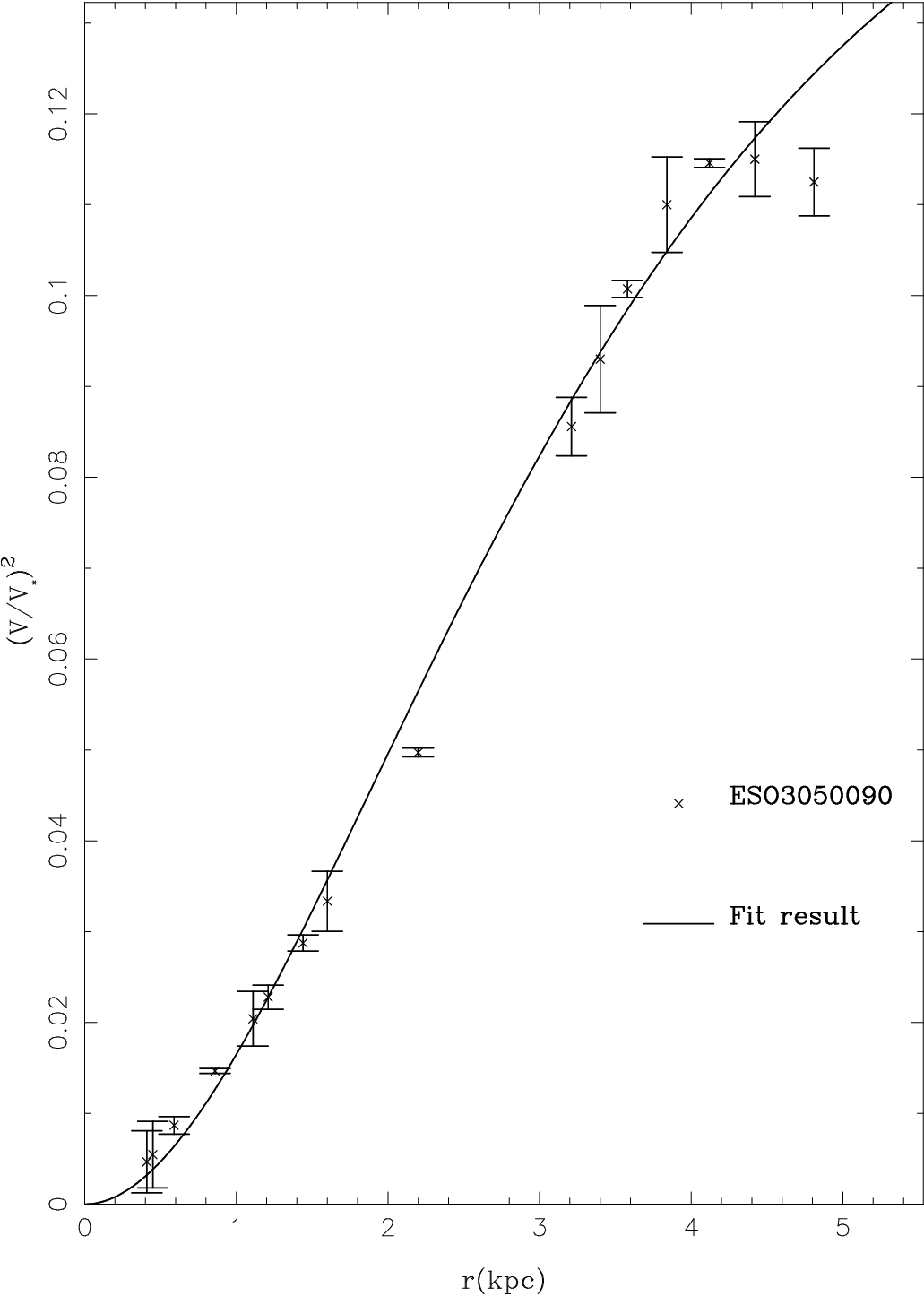} \\
\includegraphics[width=0.2\textwidth]{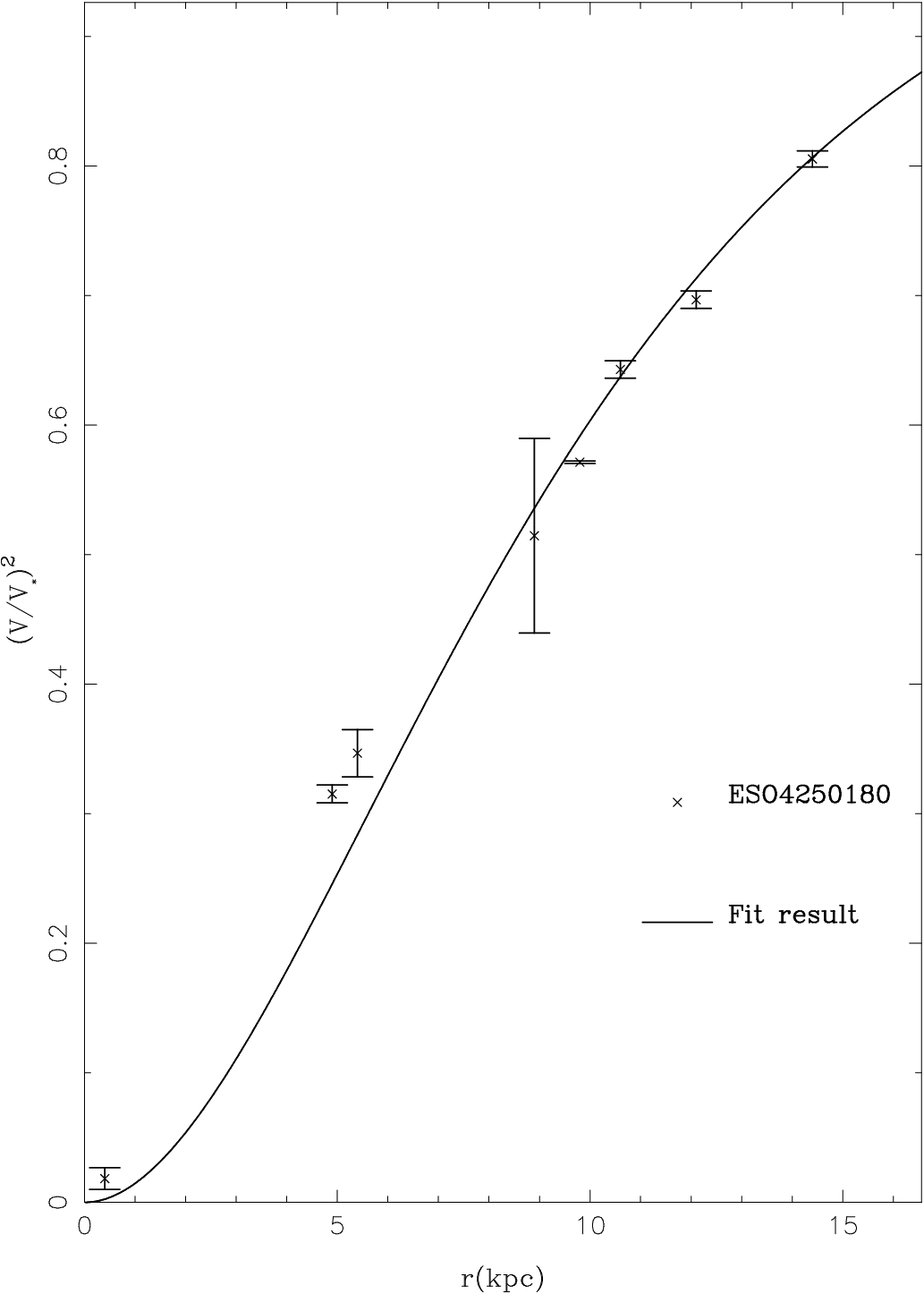}
\includegraphics[width=0.2\textwidth]{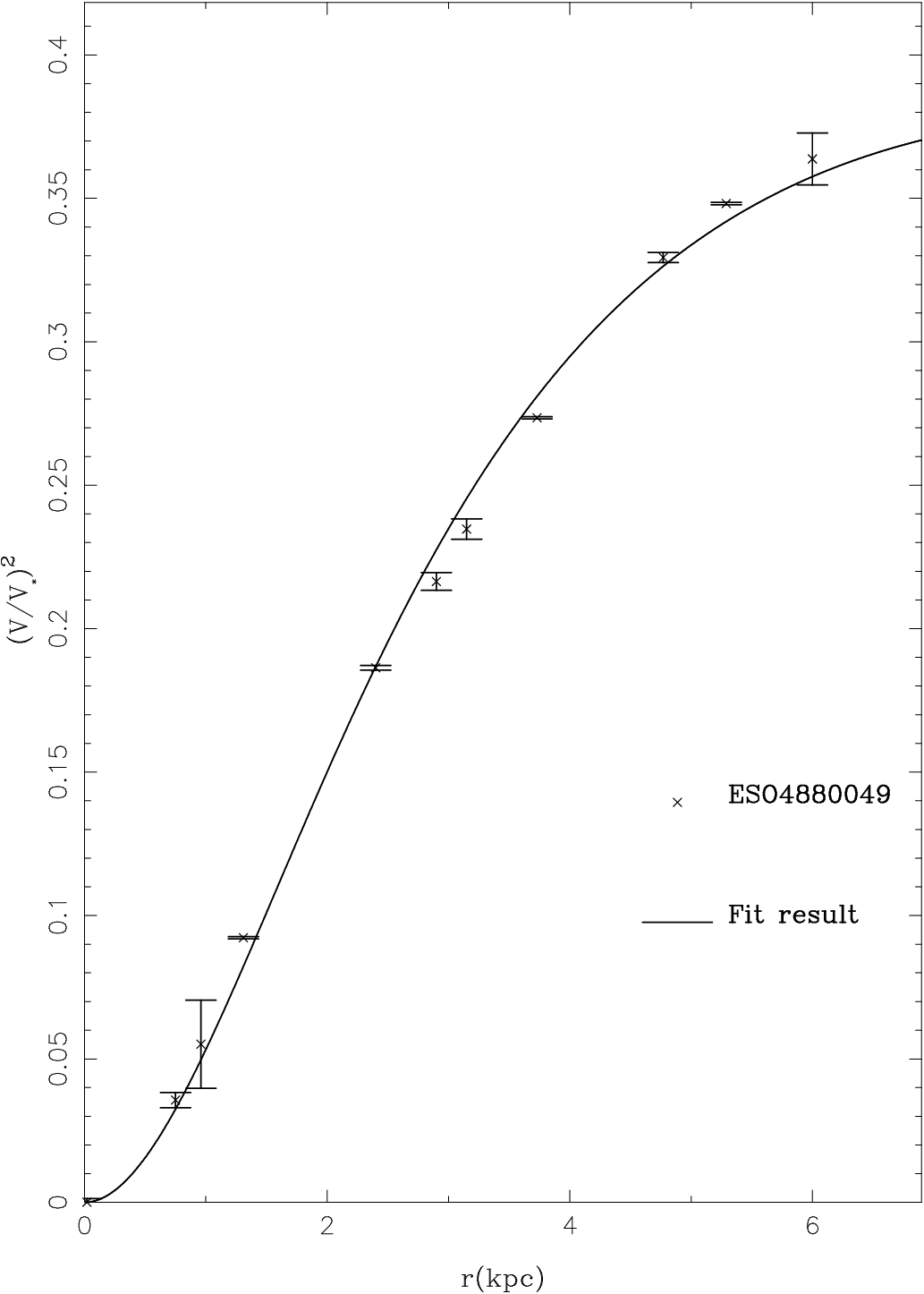}
\begin{minipage}[]{85mm}
\caption{Continued.} 
\end{minipage}
\end{figure}

\begin{figure}
\centering
\includegraphics[width=65mm]{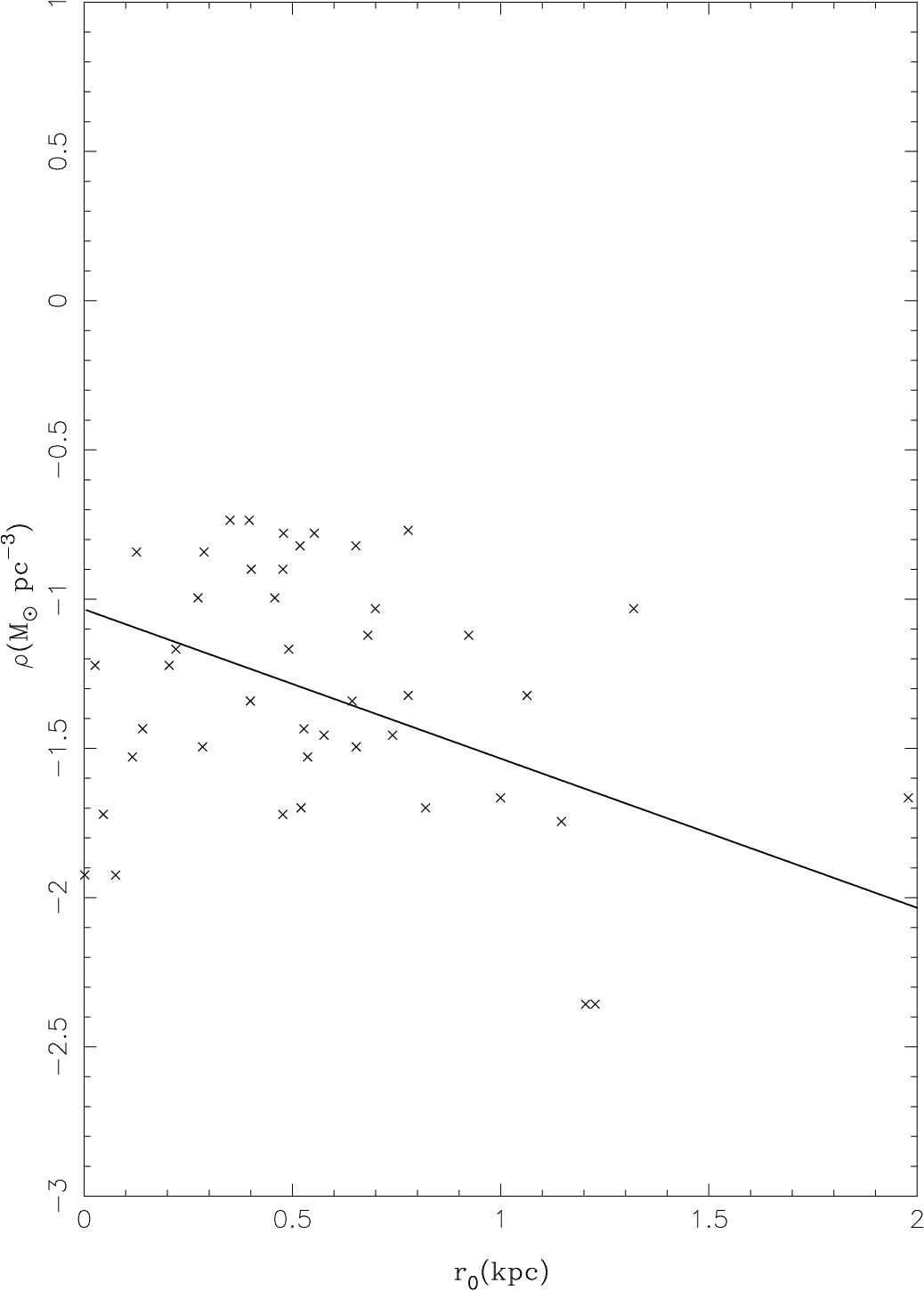} 
\includegraphics[width=63mm]{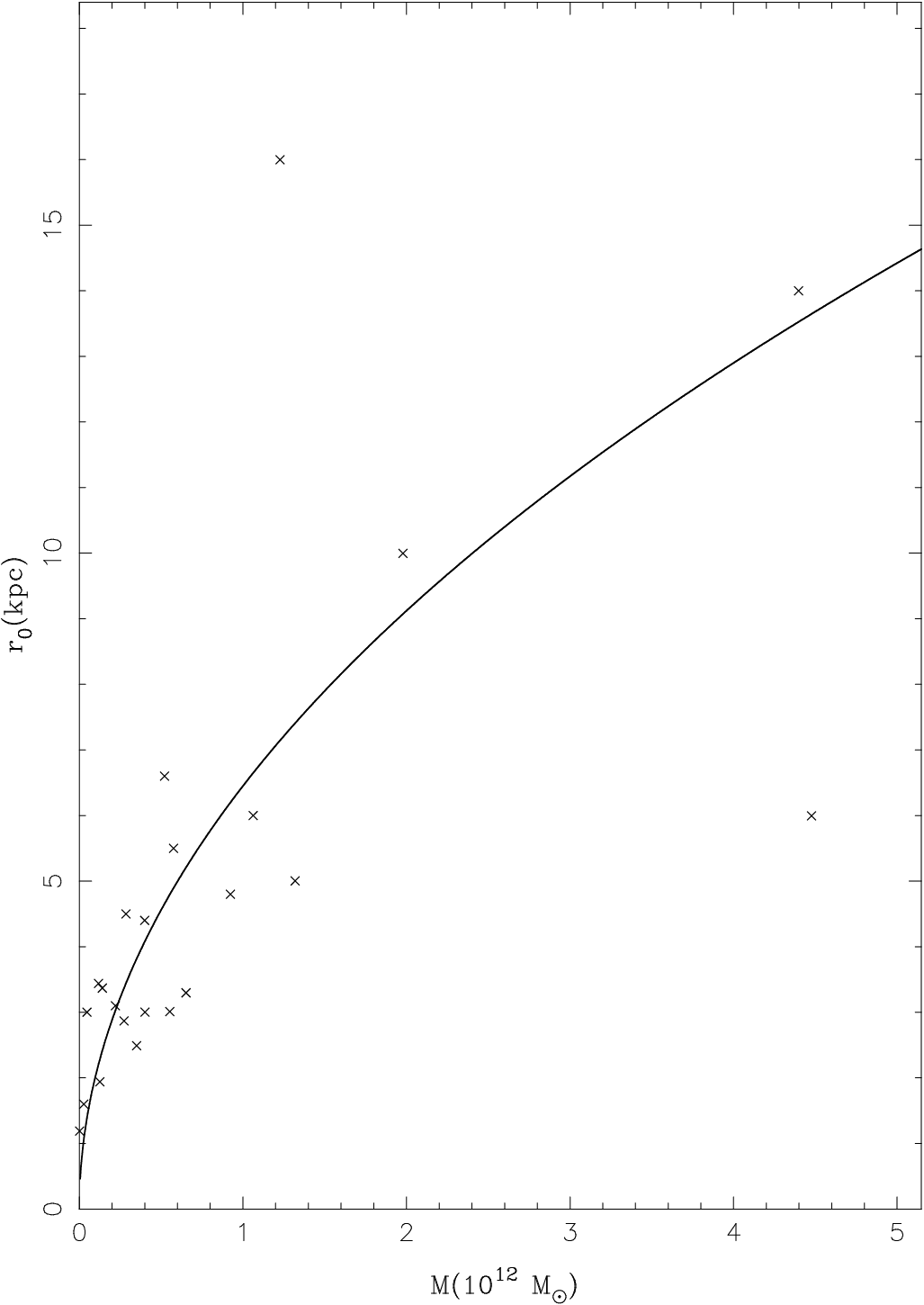}
\caption{Left panel: the correlation between central density $\rho_0$ and core radius $r_0$; crosses represent the data derived from the rotation curves of the sample with the Burkert profile, the solid line is the fit with the data. Right panel: the relation between core radius $r_0$ and the virialized halo mass $M$.} 
\label{rhor0M}
\end{figure}

The corresponding orbital velocity at radius $r$ is simply $V^2(r)=GM(r)/r$, where $G$ is the gravitational constant. The sample \citep{deBlok2001a} provides for each galaxy the observed velocity $V$/(Km/s) and uncertainty $V_{Err}$/(Km/s) at radius $r$/Kpc. We fit the rotation curves with $V^2(r,\rho_0,r_0)$ and derive the values of the parameter pair ($\rho_0$, $r_0$) for each galaxy of the sample. Figure \ref{Vr} shows the fitted $V^2(r)$ (solid lines) and the observed data of rotation curves (crosses) for each galaxy. Note that, for a few galaxies, especially F579, U11648 and U11748, the Burkert-fitted lines (solid) deviate greatly from the data points at large radii. However, this will not affect our further analysis, because our interests are focused on the central regions, where the data have been well-fitted. Statistically, there is a correlation between $\rho_0$ and $r_0$ for the sample; the fitted result is  
\begin{equation}
 \rho_0=0.029\left(\frac{r_0}{\textrm{Kpc}}\right)^{-1/2}M_{\sun}/\textrm{pc}^{3}.
\label{rhor0}
\end{equation}
In later lens probability caculations, we need to know the total virialized halo mass $M=M_{vir}$. For the Burkert profile, the mass diverges logarithmically at large radii, and should be cut off at $r_{200}$, the radius of a sphere within which the average mass density is 200 times the critical density of the universe, typically taken as the virial radius \citep{NFW97}. So we define 
\begin{equation}
 M\equiv M_{vir}=M_{200}=M(r_{200})=\frac{800\pi\rho_\mathrm{crit}(z)}{3}r_{200}^3,
\label{M}
\end{equation}
where $M(r_{200})$ is the virialized halo mass obtained by substituting $r=r_{200}$
into equation (\ref{Mr}), and $\rho_\mathrm{crit}(z)$ is the critical density of the universe at redshift $z$. By numerically solving Equations (\ref{rhor0}) and (\ref{M}), we get an approximate relation between $r_0$ and $M$,
\begin{equation}
 r_0=6.45\left(\frac{M}{10^{12}M_{\sun}}\right)^{1/2}\textrm{Kpc},
\end{equation}
which is similar to the result of \citet{Salucci2007}, although we have adopted different procedures and used different samples. The correlations between the pairs $(\rho_0, r_0)$ and $(r_0, M)$ are presented in Figure \ref{rhor0M}.

\section{Strong lensing probabilities}
\label{sect:lens}

It has been explicitly pointed out \citep{Mashchenko2006} that,  on one hand,  the resultant cored dark matter halos produced by stellar feedback are in agreement with the Burkert profile, and on the other hand, as the dwarf galaxies merge together to make larger ones, the flat-cored shape of the dark matter density profile is preserved in subsequent massive halos. Therefore, in this section, we assume that the Burkert profile constrained from the rotation curves of the LSB galaxies  in the previous section is also appropriate for strong lensing galaxies (usually giant ellipticals), and we use it to calculate the corresponding lensing frequency. 

Gravitational lensing provides a powerful and independent tool to detect the matter distribution in LSS \citep{Schneider1992,Wambsganss1995,Wu1996,Wu2000,Wu2004,Bartelmann1996,Bartelmann2001,Keeton2001a,Keeton2002,Keeton2003,Keeton2004,Zhang2003,Zhang2009}. In particular, by virtue of eliminating the degeneracy of some lens and cosmological models, strong lensing statistics exhibits an exceptional ability to investigate both dark matter \citep{Turner1984,Wu1993,Chae2002,Li2002,Li2003,Chae2003,Chen2003a,Chen2003b,
Chen2005,Chen2006,Chen2008a,Oguri2002,Oguri2003a,Oguri2003b,Zhang2004,Chae2007} and dark energy \citep{Fukugita1990,Fukugita1991,Turner1990,Krauss1992,Maoz1993,Wu1993,Kochanek1995,Kochanek1996,Falco1998,Cooray1999,Waga1999,Sarbu2001,Chen2004a,Chen2004b,Dev2004,Yang2009,ZhangQJ2009}. With the increasing number of the newly discovered lenses \citep[e.g.,][]{Wen2009}, we now have at least two well-defined statistical lens samples. In addition to the radio lens sample \cite[CLASS/JVAS ;][]{Patnaik1992,King1999,Myers2003,Browne2003}, we  have a Sloan Digital Sky Survey Quasar Lens Search Data Release 3 \citep[SQLS DR3;][]{Inada2008} optical lens sample, which first included a cluster-scale lens, and thus provides stronger constraints on halo models, especially for the universal properties of halos ranging from dwarf galaxies to clusters of galaxies \citep{Oguri2008,Oguri2009}.

The lens equation is
$\eta=D_\mathrm{S}\xi/D_\mathrm{L}-D_\mathrm{LS}\hat{\alpha}$,
where $\eta$ and $\xi$ are the physical positions of a source in
the source plane and an image in the image plane, respectively,
$\hat{\alpha}$ is the deflection angle, and $D_\mathrm{L}$,
$D_\mathrm{S}$, and $D_\mathrm{LS}$ are the angular diameter
distances from observer to lens, observer to source, and lens to
source, respectively. By defining dimensionless positions
$y=D_\mathrm{L}\eta/D_\mathrm{S}r_0$ and $x=\xi/r_0$, and
dimensionless angle
$\alpha=D_\mathrm{L}D_\mathrm{LS}\hat{\alpha}/D_\mathrm{S}r_0$,
the lens equation is then $y=x-\alpha$. For circularly symmetric lens, 
$\hat{\alpha}=4GM(\xi)/c^2\xi$, where $M(\xi)$ is the mass within a circle of radius $\xi$ \citep{Schneider1992}. For the Burkert profile, $M(x)=4\pi\rho_0r_0^3m(x)$ \citep{Park2003}, where
\begin{displaymath}
 m(x)=\left\lbrace \begin{array}{ll} 
\ln\frac{x}{2}+\frac{\pi}{4}(\sqrt{x^{2}+1}-1)+
\frac{\sqrt{x^{2}+1}}{2}\textrm{arccoth}(\sqrt{x^{2}+1}) \\
-\frac{1}{2}\sqrt{x^{2}-1}\arctan\sqrt{x^{2}-1}, & \textrm{if  $x>1$ }, \\
& \\
-\ln2-\frac{\pi}{4}+\frac{1}{2\sqrt{2}}[\pi+\ln(3+2\sqrt{2})], & \textrm{if  $x=1$}, \\
& \\
\ln\frac{x}{2}+\frac{\pi}{4}(\sqrt{x^{2}+1}-1)+\frac{\sqrt{x^{2}+1}}{2}\textrm{arccoth}(\sqrt{x^{2}+1})\\
+\frac{1}{2}\sqrt{1-x^{2}}\textrm{arctanh}\sqrt{1-x^{2}}, & \textrm{if  $x<1$}.\end{array} \right.
\end{displaymath}
The lens equation for the Burkert profile then is
\begin{equation}
 y=x-\frac{8\kappa_c}{\pi}\frac{m(x)}{x}, \ \ \ \ \kappa_c
=\left(\frac{2\pi^2G\rho_0r_0}{c^2}\right)\left(\frac{D_\mathrm{L}D_\mathrm{LS}}{D_\mathrm{S}}\right)
=\frac{\Sigma(0)}{\Sigma_\textrm{crit}},
\label{lenseq}
\end{equation}
where $\kappa_c$ is the central convergence, an important quantity that determines wheither strong lensing can occur or not, $\Sigma(0)=\pi\rho_0r_0/2$ is the surface mass density at the lens center and $\Sigma_\textrm{crit}=c^2D_\mathrm{S}/4\pi GD_\mathrm{L}D_\mathrm{LS}$ is the critical surface density.

Generally, for any spherically symmetric
density profiles of lensing halos, multiple images can be produced
only if the central convergence $\kappa_c$ is greater than unity
\citep{Schneider1992}. When $\kappa_c\leq 1$, only one image is produced.
Note that even if $\kappa_c>1$ is satisfied, multiple images can
occur only when the source is located within
$y_\mathrm{cr}=y(x_\mathrm{cr})$ \citep{Li2002}, where $x_\mathrm{cr}$ is
determined from equation(\ref{lenseq}), with 
$dy/dx=0$ for $x<0$ \citep[this is similar to lensing by cored isothermal sphere halos, see][]{Chen2005}. For a singular
density profile such as the singular isothermal sphere (SIS) and NFW profiles, the central value 
is divergent, so $\kappa>1$ is always satisfied, and multiple images can be
produced for any given mass. For density profiles with a finite
soft core such as core isothermal sphere, NTIS and Burkert profiles, however, the condition 
$\kappa>1$ requires that only halos with mass greater than a certain value (determined
by $\kappa_c=1$) can produce multiple images.

\begin{figure}[!ht]
\centering
 \includegraphics[width=9.0cm, angle=0]{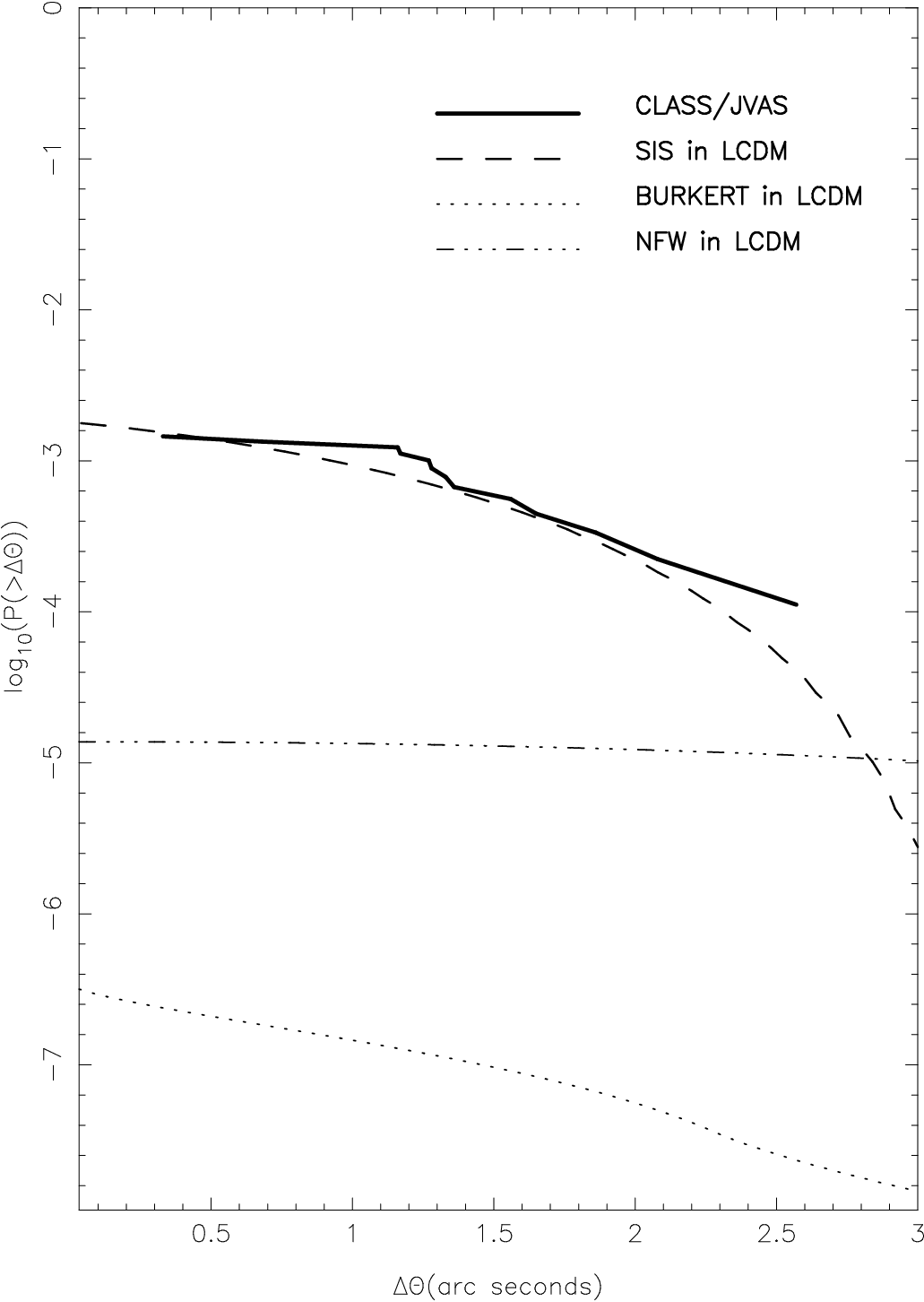}
\caption{Lensing probability with image separation larger than $\Delta\theta$. Our predicted lensing probability based on the Burkert model (\textit{dotted line}) is about four orders of magnitude lower than the observations of CLASS/JVAS (\textit{thick hostogram}). As expected, the SIS model (\textit{dashed line}) is in veru good agreement with observations. The NFW model (\textit{dash-dotted}) also fails to explain observations but is much better than the Burkert model.}
\label{Figprob}
\end{figure}

The calculations for lensing probabilities with quasars at redshift $z_s$ lensed by Burkert halos of galaxies and with image separations larger than $\Delta\theta$ are straightforward \citep[see][for details]{Chen2005},
\begin{equation}
 P(>\Delta\theta)=0.06y_\mathrm{cr}\int_{0}^{1.27}(1+z)^{3}\frac{dD_{L}^{P}(z)}{dz}dz \int_{M_\mathrm{min}(\Delta\theta,z)}^{\infty}f(M,z) dM\int_{0}^{y_\mathrm{cr}}y[\mu(y)]^{1.1}dy,
\label{lensprob}
\end{equation}
where $D_{L}^{P}(z)$ is the proper distance from the observer to the lens located at redshift $z$; $f(M,z)$ is the mass function, for which we use the expression given by \citet{Jenkins2001}; $M_{\mathrm{min}}(\Delta\theta,z)$ is the
minimum mass of halos above which lenses can produce images with
separations greater than $\Delta\theta$; and $\mu(y)$ is the total
magnification of the two outer images for a source at $y$, which can be calculated directly from the lens equation (\ref{lenseq}). 

In our actual numerical calculations, we choose almost the same procedures, mass function and even background universe as in \citet{Chen2005},  except for the distribution of the redshifts of  source quasars, for which, as in \citet{Chen2008}, we use only the mean value $z_s=1.27$. The result is shown in Figure \ref{Figprob}. The observational result for the well defined combined sample of the cosmic Lens All-Sky Survey (CLASS) and Jodrell Bank/Very Large Array Astrometric Survey (JVAS) is shown as thick histogram \citep[see][and references therein]{Chen2005}. For comparison, the lensing probability of the SIS model and NFW model are also shown with the same parameters and approximations as Burkert.

\section{Discussion and conclusions}
\label{sec:discussion}

We have calculated the lensing probability based on the Burkert profile model, for which the core size-halo mass relation is obtained by fitting the high-resolution rotation curves of a sample containing 26 galaxies. For comparison, we have recalculated the lensing probability for the SIS model and NFW model with the same parameters and approximations as Burkert. As expected, the lensing probability for the Burkert model is 4 orders of magnitude lower than the observations of CLAS/JVAS. While our stong lensing calculations rule out the Burkert model as a possible giant elliptical density profile, we conclude, at the same time, that the sellar feedback mechanisms \citep{Mashchenko2006,Peirani2008} are also ruled out as reasonable solutions to the cusp-core problem. Additionally this conclusion is  independent of the models used to fit rotation curves: Burkert in this paper, and cored isothermal sphere \citep{deNaray2008} in \citet{Chen2008}. 

Other solutions to the cusp-core problem, including the analytical NTIS model \citep{Shapiro1999,Iliev2001,Chen2005}, dynamical friction of substructures \citep{El-Zant2001,Tonini2006,Romano2008} and   stellar bar-CDM interaction \citep{Weinberg2002,Holley2005}, face exactly the same embarrassing problem. Clearly, the solutions proposed so far can produce a large core for each dwarf and LSB galaxy, and thus can successfully explain the observations of rotation curves, but they cannot explain the steep and cuspy centers of massive galaxies, which are favored by stong lensing and X-ray observations. As a matter of fact, in the framework of LCDM, the so called ``cusp-core problem'' encountered by hierarchical structure formation theories is not just the contradiction between the cusp predicted by pure CDM simulations and the core implied by the observations of rotation curves, but rather, the cusp predicted by pure CDM simulations on all mass scales, the core implied by rotation curves on small, dwarf galaxy scales and the cusp favored by strong lensing and X-ray observations on large, giant elliptical galaxy scales. From an observational point of view,  regardless of theories, we can call this the  ``cusp-core phenomena'' rather than ``problem'', when referring to the cusp in large scale galaxies and core in small scale galaxies. As we have pointed out, a reasonable, self-consistent theory should explain the observations on all mass scales, i.e., all ``cusp-core phenomena'', rather than only the cores on small scales.

\normalem
\begin{acknowledgements}
We are indebted to Stacy McGaugh for providing us with detailed data of the rotation curves of the sample, and we are grateful to the anonymous referee for helpful suggestions and comments. This work was supported by the National Natural Science Foundation of China under grant 10673012, CAS under grant KJCX3-SYW-N2 and the National Basic Research Program of China (973 Program) under grant 2009CB24901.
\end{acknowledgements}

\label{lastpage}

\end{document}